\documentclass{article}
\usepackage[big]{dgruyter}

\usepackage[utf8]{inputenc}
\usepackage{microtype}
\usepackage{gensymb}
\usepackage{natbib}
\usepackage{csquotes}
\usepackage{hyperref}

\usepackage{graphicx}
\usepackage[caption=false]{subfig}

\usepackage{siunitx}
\usepackage{textcomp}
\usepackage{amsmath}
\usepackage{amssymb}

\usepackage{acronym}
\newacro{radial}[R]{radial}
\newacro{tangential}[T]{tangential}
\newacro{longitudinal}[L]{longitudinal}
\newacro{c4}[C4]{Computer Controlled Climatized Creep Rack}
\newacro{rh}[RH]{relative humidity}
\newacro{dic}[DIC]{digital image correlation}
\newacro{ni}[NI]{National Instruments}
\newacro{roi}[ROI]{region of interest}

\begin{document}

\articletype{Research Article{\hfill}Open Access}
    \author[1]{Jonas M. Maas}
    \author[2]{Falk K. Wittel}
    
    \affil[1]{ETH Zurich, Institute for Building Materials, HIF E 27, Laura-Hezner-Weg 7, CH-8093 Zurich, Switzerland; E-mail: jomaas@ethz.ch; ORCID: 0000-0001-5679-7352}
    \affil[2]{ETH Zurich, Institute for Building Materials, HIF E 27, Laura-Hezner-Weg 7, CH-8093 Zurich, Switzerland; E-mail:  fwittel@ethz.ch; ORCID: 0000-0001-8672-5464}

    \title{\huge Comprehensive creep compliance characterization of orthotropic materials using a cost-effective automated system}
    \runningtitle{Creep compliance characterization of orthotropic materials}
    
	\begin{abstract}
		{
            Determining the creep compliances of orthotropic composite materials requires experiments in at least three different uniaxial and biaxial loading directions. Up to date, data respecting multiple climates and all anatomical directions are sparse for hygro-responsive materials like Norway spruce. Consequently, simulation models of wood frequently over-simplify creep, e.g., by proportionally scaling missing components or neglecting climatic influences.
            To overcome such simplifications, an automated computer-controlled climatized creep rack was developed, that experimentally assesses moisture-dependent viscoelasticity and mechanosorption in all anatomical directions. The device simultaneously measures the creep strains of three dogbone tension samples, three flat compression samples, and six Arcan shear samples via Digital Image Correlation. This allows for ascertaining the complete orthotropic compliance tensors while accounting for loading direction asymmetries.
            This paper explains the creep rack's structure and demonstrates its use by determining all nine independent creep compliance components of Norway spruce at 65\% relative humidity. The data shows that loading asymmetry effects amount up to 16\%. Furthermore, the found creep compliance tensor is not proportional to the elastic compliance tensor. By clustering the compliance components, we identify four necessary components to represent the full orthotropy of the compliance tensor, obtainable from not less than two experiments.
        }
	\end{abstract}

    \keywords{viscoelasticity; experimental; automatization; Norway spruce}

    \startpage{1}
    
    \journalyear{2024}

\maketitle


\section{Introduction} 
Predicting the rheological behavior of a wood species necessitates accurate simulation models founded on comprehensible experimental data. This data usually depends on the moisture content, anatomical direction, and loading direction \citep{niemz2023, kollmann1968}. While elasticity and plasticity are well-researched (e.g., \citealp{neuhaus1981, schmidt2006, hering2012_1}), knowledge for moisture-dependent viscoelastic and mechanosorptive creep is sparse. Many publications investigated distinct creep properties, but rarely all nine independent components of the creep compliance tensor.

\citet{gressel1983} and \cite{tong2020} review studies on viscoelastic creep data in distinct anatomical directions. These studies provide creep data for tension or compression parallel and perpendicular to the grain, bending, and torsion. Further studies, like \citet{schniewind1972}, \citet{hayashi1993}, \citet{taniguchi2010}, and \citet{jiang2016}, investigate the relationships of axial to lateral strains. They conclude that Poisson's ratios are not constant during creep, but their observations do not consistently show whether they increase or decrease over time. \citet{hilton2001} argues from mechanical considerations that Poisson's ratios must not be constant. Yet, he recommends directly describing creep by the axial and transverse compliances because small errors in  Poisson's ratios could lead to severe distortion of the creep compliance matrix.

The studies of \citet{schniewind1972}, \citet{cariou1987}, \cite{taniguchi2010}, and \cite{ozyhar2013} determine the relative axial creep for multiple anatomical directions. They indicate more pronounced creep in the tangential and radial directions compared to the longitudinal direction.
Moreover, comparative experiments \citep{gressel1983, ozyhar2013} show that compression samples creep more than tensile samples. The loading degree additionally influences the creep response. The threshold for linear creep ranges from 30\% to 84\%, depending on the literature source, moisture content, and anatomical direction \citep{schniewind1968, gressel1983, liu1993, morlier1994}.

Numerous studies also investigated shear creep. \citet{schniewind1972} utilized plate shear tests, while \citet{hayashi1993}, \citet{bengtsson2023}, \citet{ando2023}, and \citet{shimazaki2024} employed off-axis tests. While off-axis tests have been valuable in various contexts, they have inherent limitations because they do not yield a uniform shear and normal stress state \citep{ho1993, pierron1996, xavier2004}. \cite{akter2023} determined the rolling shear in one shear plane for five different relative humidities.

The mentioned publications show that creep depends on the anatomical direction, loading type, moisture content, and loading degree. However, the data obtained from these studies is inconsistent. They mostly originate from one climate, different species, different loading cases, or limited anatomical directions. Furthermore, most studies' experimental runtime was below 24 hours, thus neglecting long-term effects. Except for \citet{cariou1987}, who investigated Maritime pine, studies determining all nine creep compliance components are rare. For mechanosorptive creep, the data is similarly fragmented, and similar effects can be observed \citep{rantamaunus1975, toratti1992, toratti2000, huc2018, niemz2023}.

The incomplete data forces developers of computational simulation models to scale available compliances between anatomical directions, trees, humidities, and wood species  \citep{hanhijarvi2003, fortino2009, hassani2015, yu2022}. The scaling ultimately leads to severe inaccuracies in model responses. Experimental parameters for every component of the moisture-dependent viscoelastic and mechanosorptive creep compliance tensor could resolve such inaccuracies -- if the parameters originate from one tree of one species. However, such a task is challenging for experimental scientists. It requires repeated time-intensive experiments in at least three uniaxial and shear directions for various climates, ideally respecting asymmetries between compression and tension and the different shear planes.

We developed a fully automatized creep rack to reduce the time expenditure of such an experimental campaign. The rack measures moisture-dependent creep deformations of three compression, three tension, and six shear samples in parallel. It includes a climate chamber, allowing for recording viscoelastic and mechanosorptive creep strains under controlled humidity conditions. This setup enables the direct determination of each component of the creep compliance tensor from one set of experimental data. Compared to commercial systems, e.g., the Zwick Kappa-Multistation \citep{zwickroell2023}, our system features lower costs (< 13.000\.Euro material costs), a higher number of testing axes, high adaptability of the components, more flexibility in the arrangement of sample shapes, and an optimization on wood testing.

In this paper, we describe the automatized creep rack and its sample design. The paper contains structural details for load application, climate control, and strain recording. As an application study of the creep rack, we determine Norway spruce's axial, lateral, and shear creep strains to assemble the full viscoelastic creep tensor at 65\% \ac{rh}. The study additionally analyzes asymmetries and proportionalities between the individual creep compliance components. By clustering these components into coincident groups, we can recommend the minimum number of experiments required to represent sufficiently the orthotropic relations in the creep compliance tensor.

\section{Materials and methods} \label{sec_methods}

\begin{figure}[!t]
    \centering
    \includegraphics[width=0.74\textwidth]{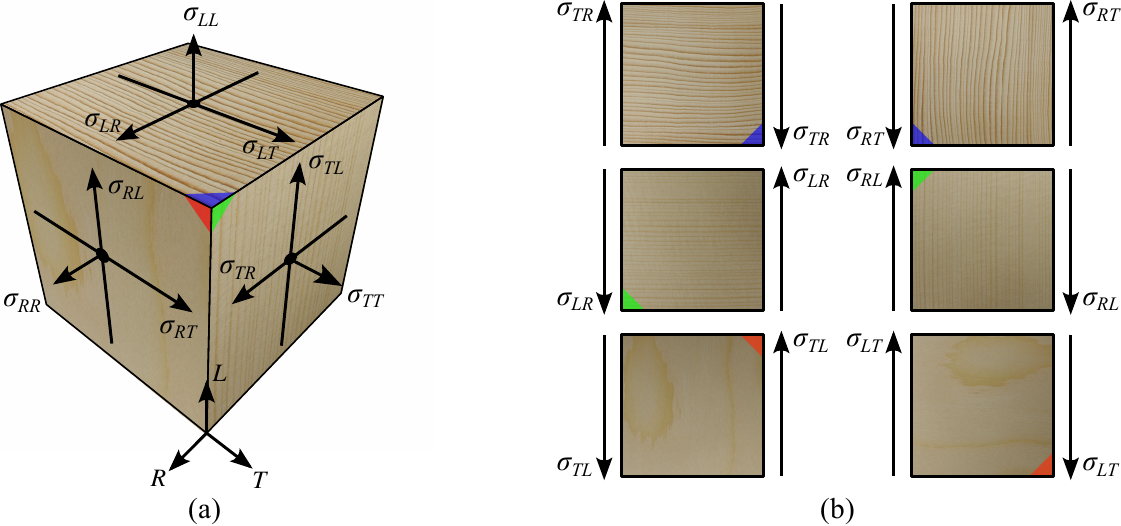}
    \caption{Mechanical planes of bulk wood in radial (R), tangential (T), and longitudinal (L) directions. $\sigma_{ii}$ are the normal stresses in direction $i$ and $\sigma_{ij}$ ($i \neq j$) the shear stresses in direction $j$ on the plane normal to $i$. (a) shows the 3D normal and shear stresses and (b) the 2D shear planes.}
    \label{fig_anatomical_layers}
\end{figure}

The goal of the creep rack is to determine every component of the time-dependent compliance tensor $\boldsymbol{C}^{-1} (t, \omega, \boldsymbol{d})$. The tensor varies with the moisture content $\omega$ and loading degree $\boldsymbol{d}$. By assuming wood as an orthotropic material \citep{kollmann1968} and omitting $\omega$ and $\boldsymbol{d}$ in the notation, the viscoelastic creep compliance $\boldsymbol{C}^{-1}(t)$ has the following form for a constant moisture content and loading degree:
\begin{align} \label{eq_creep_compliance_tensor}
\boldsymbol{C}^{-1} (t) = 
\begin{bmatrix}
C^{-1}_{11}(t) & C^{-1}_{12}(t) & C^{-1}_{13}(t) & 0 & 0 & 0 \\
C^{-1}_{21}(t) & C^{-1}_{22}(t) & C^{-1}_{23}(t) & 0 & 0 & 0 \\
C^{-1}_{31}(t) & C^{-1}_{32}(t) & C^{-1}_{33}(t) & 0 & 0 & 0 \\
0 & 0 & 0 & C^{-1}_{44}(t) & 0 & 0 \\
0 & 0 & 0 & 0 & C^{-1}_{55}(t) & 0 \\
0 & 0 & 0 & 0 & 0 & C^{-1}_{66}(t) \\
\end{bmatrix} .
\end{align}
Based on orthotropic linear elasticity \citep{kollmann1968, kienzler2019}, the diagonal components of the tensor are
\begin{align}
    \begin{aligned}
        C_{11}^{-1}(t) &= E_{R}^{-1}(t), & \quad C_{22}^{-1}(t) &= E_{T}^{-1}(t), & \quad C_{33}^{-1}(t) &= E_{T}^{-1}(t), \\
        C_{44}^{-1}(t) &= G_{RT}^{-1}(t), & \quad C_{55}^{-1}(t) &= G_{RL}^{-1}(t), & \quad C_{66}^{-1}(t) &= G_{TL}^{-1}(t), \\
    \end{aligned}
\end{align}
and the off-diagonal components are
\begin{align}
    \begin{aligned}
        C_{12}^{-1}(t) &= -\nu_{TR}(t) E_{T}^{-1}(t), & \quad C_{13}^{-1}(t) &= -\nu_{LR}(t) E_{L}^{-1}(t), & \quad C_{23}^{-1}(t) &= -\nu_{LT}(t) E_{L}^{-1}(t), \\
        C_{21}^{-1}(t) &= -\nu_{RT}(t) E_{R}^{-1}(t), & \quad C_{31}^{-1}(t) &= -\nu_{RL}(t) E_{R}^{-1}(t), & \quad C_{32}^{-1}(t) &= -\nu_{TL}(t) E_{T}^{-1}(t). \\
    \end{aligned}
\end{align}
$t$ is the time, $E_i$ is the Young's modulus in direction $i$, $G_{ij}$ is the shear modulus for the shear plane normal to $i$ in force direction $j$, and $\nu_{ij}$ is the Poisson's ratio with axial direction $i$ and lateral direction $j$. $i$ and $j$ match the anatomical directions and are radial (R), tangential (T), and longitudinal (L) to the fiber direction, as depicted in Fig.\,\ref{fig_anatomical_layers}a.

Due to the orthotropy condition $\nu_{ij}(t) E_i^{-1}(t) = \nu_{ji}(t) E_j^{-1}(t)$, Eq.\,\ref{eq_creep_compliance_tensor} is symmetric, i.e., $C^{-1}_{ij} = C^{-1}_{ji}$ for $i \neq j$. That assumption leads to nine independent creep compliance components. Determining these components requires at least three uniaxial and three shear creep experiments. The axial strains of the uniaxial creep tests quantify the following diagonal components:
\begin{align}
    \begin{aligned} \label{eq_creep_compliance_tensor_exp_axial}
        C_{11}^{-1}(t) &= \frac{\epsilon_{RR}^{RR}(t)}{\sigma_{RR}}, & \quad C_{22}^{-1}(t) &= \frac{\epsilon_{TT}^{TT}(t)}{\sigma_{TT}}, & \quad C_{33}^{-1}(t) &= \frac{\epsilon_{LL}^{LL}(t)}{\sigma_{LL}}. \\
    \end{aligned}
\end{align}
$\epsilon_{ii}^{jj}$ is the time-dependent strain in direction $ii$ induced by a uniaxial force in direction $jj$. $\sigma_{jj}$ is the corresponding uniaxial stress. The uniaxial tests' lateral strains ($i \neq j$) correspond to the compliance's off-axis components:
\begin{align}
    \begin{aligned} \label{eq_creep_compliance_tensor_exp_lateral}
        C_{12}^{-1} = \frac{\epsilon_{RR}^{TT}(t)}{\sigma_{TT}} &= C_{21}^{-1}(t) = \frac{\epsilon_{TT}^{RR}(t)}{\sigma_{RR}}, \\
        C_{13}^{-1} = \frac{\epsilon_{RR}^{LL}(t)}{\sigma_{LL}} &= C_{31}^{-1}(t) = \frac{\epsilon_{LL}^{RR}(t)}{\sigma_{RR}}, \\
        C_{23}^{-1} = \frac{\epsilon_{TT}^{LL}(t)}{\sigma_{LL}} &= C_{32}^{-1}(t) = \frac{\epsilon_{LL}^{TT}(t)}{\sigma_{TT}}. \\
    \end{aligned}
\end{align}
The three shear experiments specify the shear components of the compliance tensor:
\begin{align}
    \begin{aligned} \label{eq_creep_compliance_tensor_exp_shear}
        C_{44}^{-1}(t) &= \frac{\epsilon_{RT}^{RT}(t)}{\sigma_{RT}}, & \quad C_{55}^{-1}(t) &= \frac{\epsilon_{RL}^{RL}(t)}{\sigma_{RL}}, & \quad C_{66}^{-1}(t) &= \frac{\epsilon_{TL}^{TL}(t)}{\sigma_{TL}}. \\
    \end{aligned}
\end{align}
$\epsilon_{ij}^{ij}$ is the shear creep in the shear plane normal to direction $i$ induced by a shearing force in direction $j$. $\sigma_{ij}$ is the applied constant shear stress.
As creep experiences asymmetries between compression and tension \citep{gressel1983, ozyhar2013, bachtiar2017}, the uniaxial creep compliances must be ideally obtained from both compression and tension tests. Similarly, the shear modulus $G_{ij}$ may differ from the $G_{ji}$, because the shear forces operate in different orientations on the fibers, as illustrated in Fig.\,\ref{fig_anatomical_layers}b. Thus, the complete determination of the tensor in Eq.~\ref{eq_creep_compliance_tensor} requires twelve tests when respecting loading direction asymmetries.

\begin{figure}[!bt]
    \centering
    \includegraphics[width=\textwidth]{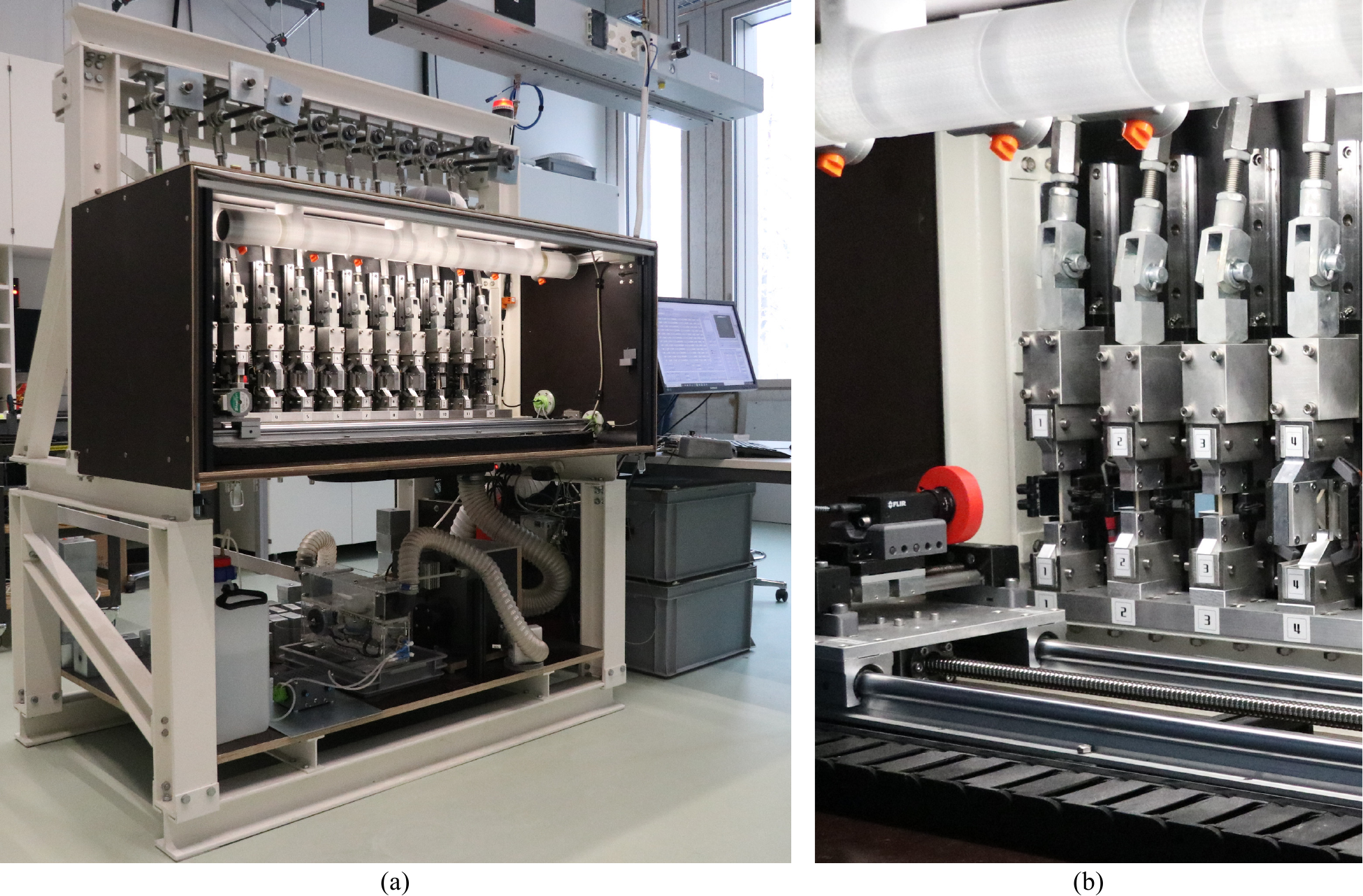}
    \caption{Frontal view of the \acf{c4}. (a) shows the supporting structure with the climate chamber that contains the test specimen. Loads are applied via a lever arm system guided by linear rails. Weights are applied to the lever arms on the backside of the rack. The section below the climate chamber holds the ventilation system. (b) shows a close-up view of the samples and camera system. The camera, connected to a 2-axis positioning system, moves past the samples and acquires images of their surfaces at fixed time intervals. The mirrors connected to the sample holders enable measuring the front and back sides of the samples.}
    \label{fig_c4_device_presentation}
\end{figure}

The creep rack, shown in Fig.~\ref{fig_c4_device_presentation}, comprises these twelve tests: one compression and tension test per anatomical direction, and six shear tests covering the six different shear planes. The tests are carried out under a controlled climate, and each sample's axial, transverse, and shear strains are recorded over time. With this data, all components of the creep compliance tensor are determined, including loading direction asymmetries. 
The device consists of the following main components: A set of twelve lever arms that apply a steady upward or downward force onto a sample, a climate chamber that keeps the samples at a target relative humidity, a ventilation system that maintains a uniform climate inside the chamber, a camera that records the strain deformations of the samples, and a 2-axis positioning system that positions the camera to acquire images of the respective sample surfaces. Mirrors connected to the sample holders enable the measurement of the front and back sides of the specimen, hence increasing the accuracy of the recorded strain deformations. The subsequent section explains the sample design in detail, the design of the creep rack's mechanical structure, the control of the applied forces and climate, the image acquisition, and the data analysis procedure based on fitting Kelvin-Voigt elements through the obtained creep curves.

\subsection{Sample design and material selection}
\begin{figure}[!b]
    \centering
    \includegraphics[width=0.8\textwidth]{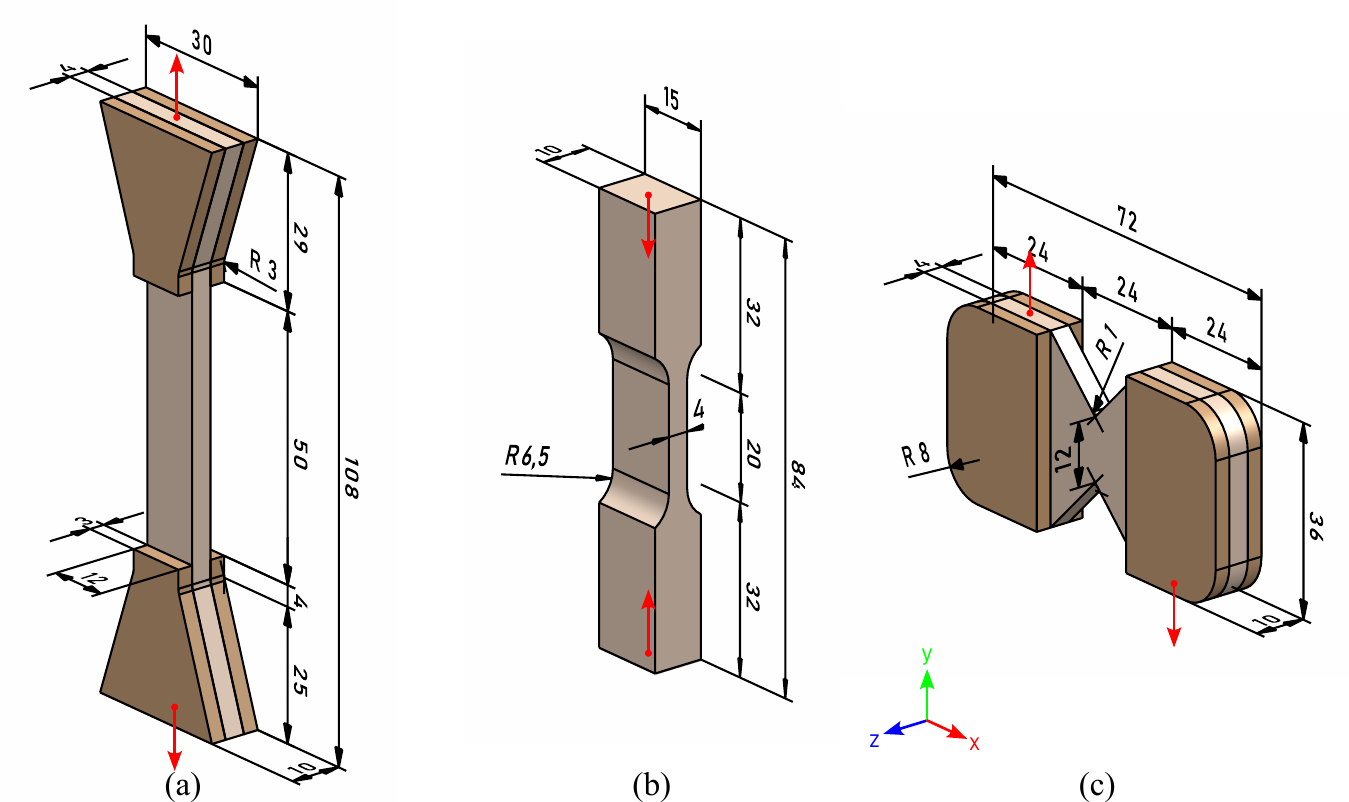}
    \caption{Sample geometries for (a) tension, (b) compression, and (c) shear creep tests. Tested bulk wood is in light brown, glued plywood strips are in dark brown (units in \si{mm}). The red arrows show the force direction. $x$, $y$, and $z$ are parallel to the wood's anatomical directions, with $y$ being the load direction and $xy$ being the shear plane.}
    \label{fig_sample_geometries}
\end{figure}

The sample geometries for the creep racks are tailored for the different loading cases like compression, tension, and shear (see Fig.~\ref{fig_sample_geometries}). The tension samples in Fig.~\ref{fig_sample_geometries}a have a flat dogbone shape, allowing for a uniform stress distribution in the center of the specimen. Their tested cross-section is $4 \times \SI{12}{mm^2}$. Glued plywood strips at the sample ends accommodate stress peaks in the mount points. The compression samples in Fig.~\ref{fig_sample_geometries}b have a comparable tested cross-section of $4 \times \SI{15}{mm^2}$ and are in a tailored shape to minimize the risk of buckling. Pre-tests in the \ac{radial} and \ac{tangential} direction, where specimens were loaded until breaking, showed perfect uniaxial failure without out-of-plane movement. The shear samples in Fig.~\ref{fig_sample_geometries}c are inspired by \citet{muller2015} and follow an Arcan-type shape, allowing for a uniform and pure shear stress distribution between the V-notches. Note that this is not the case for off-axis tests, as verified by \citet{ho1993}, \cite{pierron1996}, and \citet{xavier2004}. The cross-section between the notches is $4 \times \SI{12.4}{mm^2}$, and glued plywood strips reinforce the sample mount points. The rounded corners around the mount points are not necessarily required for such a test.

The rather small thickness of the specimens is beneficial in experiments with changing moisture levels, such as mechanosorptive tests, as they will quickly equilibrate to the climate. Thus, potential moisture gradients inside the cross-section will remain low, minimizing the internal stresses during drying and moistening. Simultaneously, the samples are wide enough to contain several growth rings along their cross-section. According to \citet{knigge1966}, the growth ring width of a Norway spruce of average density (\SI{430}{kg/m^3}) is commonly $2$ to $\SI{3}{mm}$, resulting in more than four rings along the tested cross-section width of \SI{12}{mm}. The tree investigated in this study even features growth rings of \SI{1}{mm} annual width close to the cambium. By encompassing such a sufficient number of annual rings in the specimen's cross-section, we ensure that the material parameters represent homogenized bulk wood.

As the naturally grown structure of wood leads to a large statistical variability of its mechanical properties, we recommend minimizing that variation by manufacturing all experimental samples from one reference tree. A homogeneous growth, a low number of knots, and narrow annual rings should characterize that reference tree. At the same time, the wood's density should represent the average of its species to allow for a robust scaling of the measured properties to trees of different densities. Thermo-mechanical influences from industrial manufacturing processes were avoided by harvesting a fresh tree from the forest and equilibrating it slowly to the test's relative humidity.

Even when picking a homogeneous trunk, the material behavior varies along radial and longitudinal directions \citep{trendelenburg1955, knigge1966}. Thus, to spatially backtrack every sample inside that tree, all samples used in the creep rack have a unique name comprising its cylindrical coordinates in the stem. The denotation follows the pattern \texttt{N\_TYPE-XXX-YYY-ZZZ}. \texttt{N} is the log's number if the tree got divided into multiple parts, and \texttt{TYPE} corresponds to the abbreviation of the sample type and its anatomical orientation. The cylindrical coordinates of the specimen are given by its angle relative to the log's primary cut \texttt{XXX}\,[$^\circ$], its radial distance to the pith \texttt{YYY}\,[mm], and its longitudinal distance to the log's bottom \texttt{ZZZ}\,[mm].

The tree studied in this campaign is a Norway spruce harvested in winter 2021/2022 from a Bannwald in the municipality of Alpthal, canton Schwyz, in Switzerland. It grew on \SI{1093}{m} altitude, with eastern exposition, at 47.065046°\,N, 8.710506°\,E. To minimize internal stresses and avoid cracking during drying, the tree was cut into wedge-shaped slices before air-drying it slowly to 65\% \ac{rh} at 20°C. After equilibration, we planed the wood beams to cross-sections matching the bounding boxes of the respective sample type and sliced them into the samples' outer shape. A laser cutter then cut the round corners of the shear samples. To avoid potential thermal influences on the measured mechanical properties, we ensured a minimum distance of \SI{3}{mm} between any laser cutting lines and the specimens' tested cross-section. Eventually, we milled the final geometries with a milling machine and milling frames.

Before testing, we apply a black and white airbrush speckle pattern onto the specimens' surfaces for \ac{dic}. Subsequently, we glue the plywood cap strips, laser-cut from birch plywood, with PRF (Aerodux 185 and hardener HRP 150) onto the specimens. 3D-printed gluing frames ensure an accurate orientation of the cap strips. After the adhesive bonding and re-equilibration to 65\% \ac{rh}, we mount the samples into the creep rack's sample holders, equilibrate them to the tested target climate, and then apply the weights onto the lever system. The creep rack contains eleven additional twin samples without cap strips to capture the specimens' moisture content. The moisture content is determined by weighing them at the end of each test and measuring their oven-dry mass after an experimental series.

\subsection{Load application}

\begin{figure}
    \centering
    \includegraphics[width=0.9\textwidth]{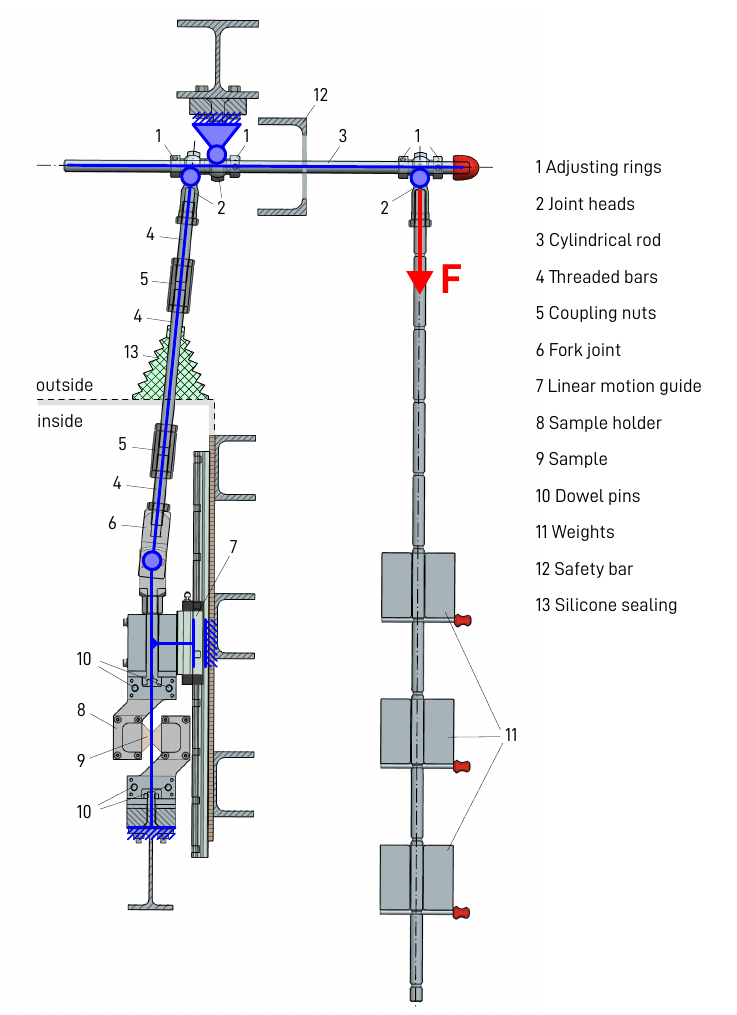}
    \caption{Structural model (blue) of the lever arm system, shown for a shear experiment. Adjustments of the horizontal location of the joint heads (2) on the cylindrical rod (3) determine the length of the lever arm and thus the relationship between the applied force $F$ and the exerted sample force. The overlap of the threaded bars (4) with the coupling nuts (5) specifies the height of the system. The linear motion guide (7) minimizes the momentum and transversal forces in the sample (9). For compression tests, the horizontal location of the left and middle joint head are interchanged. ``outside'' and ``inside'' mark the borders of the climate chamber.}
    \label{fig_lever_arm}
\end{figure}

Twelve lever systems for sample loading are mounted in parallel on a rigid frame. Rigidity is required to avoid interference between deformations of neighboring lever arms. To minimize the bending moment along the structure, the lever arms of the longitudinal experiments, exerting the highest forces, are located adjacent to the corners of the frame. In contrast, the levers for the shear experiments are positioned in the frame's center, as their applied forces are comparatively small.

Fig.~\ref{fig_lever_arm} shows the setup and structural model of the lever system with a shear sample holder. Only the sample holder differs between the test types. The lever consists of standard parts, reducing the construction costs. The relationship of weights (11) to the joint heads' (2) horizontal locations determines the applied sample load. That horizontal location is continuously adjustable along the cylindrical rod (3) and gets fixed through adjusting rings (1), allowing for a precise load calibration. By using threaded bars (4) and countered coupling nuts (5), we can change the vertical height of the lever system to the specimen's needs. The weights (11) are outside the climate chamber to keep them accessible during an experiment. The sealing (13) between the lever system and climate box consists of a deformable silicone cover. Close to the fork joint (6) is a linear motion guide (7) that avoids possible horizontal movements and rotations induced by the lever system, leaving the sample (9) under pure vertical load. Dowel pins (10) are used to mount the sample holders (8) to the lever system. Low tolerances prevent a relative rotation of the sample holders to each other, thus prohibiting rotation of the sample holders.

Before a new test setup, the loads of each lever arm get calibrated with a Burster model 8427 low-cost 10kN load cell. The load cell's maximum force relates to the failure load of a Norway spruce sample under longitudinal tension with a safety factor of $1.5$. The load cell has adapters that fit into each lever's sample holder mounts. While measuring the force of a lever, the arm's lengths and weights are adjusted to match the target load of the experiment. The callibrated cell transmits its signal to a \ac{ni} 9237 4-Channel Bridge Input Module inside a \ac{ni} 9171 USB cDAQ Chassis. A LabVIEW program interprets the bridge's digital force signal and applies a simple moving average filter with 10 seconds width to increase its accuracy.

\subsection{Climate control}

The creep rack is equipped with a moisture-controlled climate chamber made of waterproof phenolic resin boards sealed at their splices with silicone sealant. It allows for altering the experimental climates between 25\% and 90\% relative humidity.

\begin{figure}
    \centering
    \includegraphics[width=\textwidth]{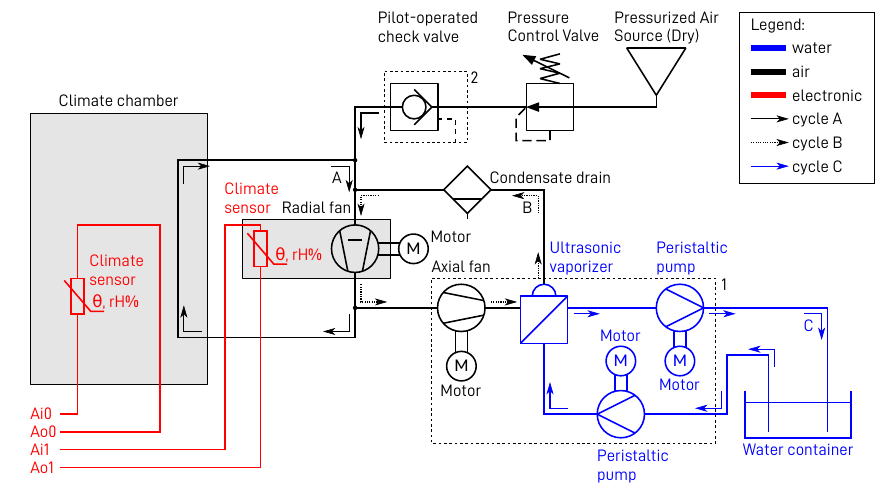}
    \caption{Climate control of the creep rack. When activated, the devices in box\,(1) increase the climate chamber's relative humidity, and the pilot-operated valve in box\,(2) decreases it. Ai and Ao are the micro-controller's analog input and output, actuating the climate sensors. A LabVIEW program interprets the climate sensors' signal and adequately controls the moistening and drying cycles.}
    \label{fig_climatecontrol}
\end{figure}

Fig.~\ref{fig_climatecontrol} illustrates the ventilation system regulating the relative humidity. The system comprises three cycles (A), (B), and (C). A radial ventilator, powered by an external motor to separate the motor's heat buildup from the airflow, drives the main air cycle (A). This cycle continuously exchanges the air volume within the climate chamber. Its 3D-printed outlet and inlet vents are located above and behind the samples, creating a constant airflow around the specimens to ensure quick acclimatization to changes in the airflow's relative humidity.

The creep rack employs humidity modules with integrated NTC temperature sensors (B+B sensors, HYTE-ANA-1735 with an accuracy of $\pm 3\%$ \ac{rh} between 20\% and 90\%\,\ac{rh} at 23°C). These sensors monitor the temperature and relative humidity inside the climate chamber and the ventilation system. They were calibrated with an Ahlborn TH climate sensor FH A646-E1 ($\pm 2\%$ \ac{rh} at $25 \pm 3$°C) and an Almemo 710 data logger.

When requiring a humidity increase, the system activates the components in Fig.~\ref{fig_climatecontrol}, box\,(1). Upon activation, the peristaltic pumps of the fluid cycle (C) supply an ultrasonic vaporizer with deionized water. The atomizer releases a damp vapor into the secondary air cycle (B), powered by an axial PC ventilator. Consecutively, cycle (B) transports the humidified air into the primary air cycle (A), raising the relative humidity within the climate chamber. A condensate drain catches excess liquid water from that cycle. Conversely, when the chamber requires a humidity decrease, the pilot-operated check valve in Fig.~\ref{fig_climatecontrol}, box\,(2) opens. This releases dry, pressurized air with less than 10\% RH from the laboratory supply into the primary ventilation system (A), thus reducing the relative humidity in the chamber.

\begin{figure}
    \centering
    \includegraphics[width=\textwidth]{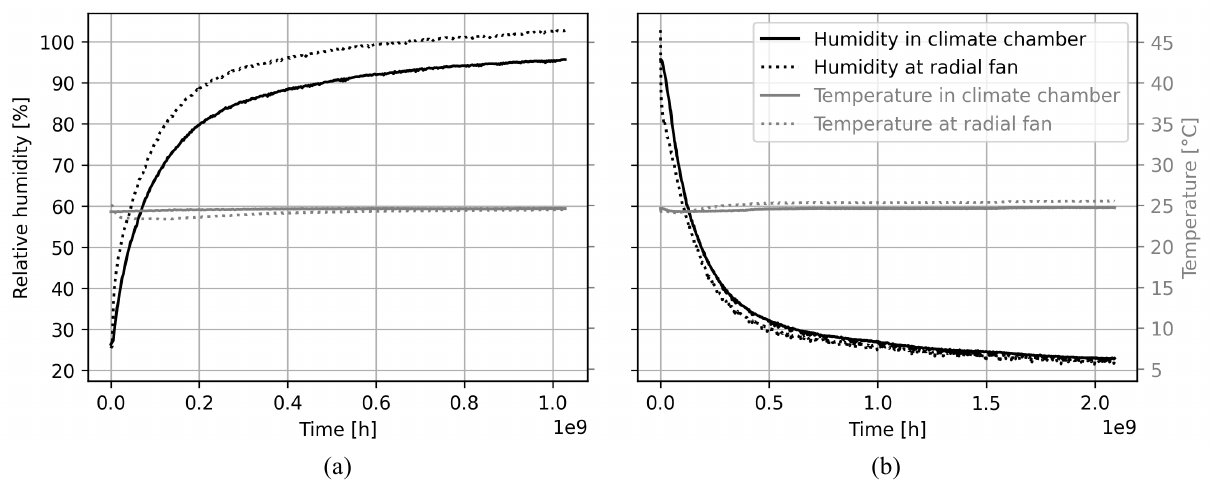}
    \caption{Moisture dynamics of the climate chamber. (a) shows the moistening curve from 26\%\,RH to 96\%\,RH and (b) the drying curve from 96\%\,RH to 23\%\,RH.}
    \label{fig_moisturedynamics}
\end{figure}

A RedLab\,1208FS USB DAQ Minilab pilots the climate control's components. The Minilab features analog and digital input/output ports, an SSR-D32A380/S relay, and accessibility by LabVIEW. The experimental control, implemented as a LabVIEW program, processes the sensor's climate signal and triggers the required ventilation system components to reach the experiment’s target climate.

The large air volume within the climate chamber affects its rate of humidity change during experiments with alternating moisture. Fig.~\ref{fig_moisturedynamics} depicts the chamber's drying and moistening dynamics. In Fig.~\ref{fig_moisturedynamics}a, the moistening dynamics show an increase from 26\%\ac{rh} to 96\%\,\ac{rh}. The relative humidity reaches 83\% after \SI{15}{min}, 90\% after \SI{30}{min}, and 95\% after \SI{60}{min}. In comparison, the drying dynamics in Fig.~\ref{fig_moisturedynamics}b are slower. When drying from 96\%\,\ac{rh} to 23\%\,\ac{rh}, the relative humidity reaches 43\% after \SI{15}{min}, 32\% after \SI{30}{min}, 27\% after \SI{60}{min}, and 23\% after \SI{120}{min}. The humidity in air cycle (A), next to the main radial fan, changes more rapidly than in the chamber due to the injection of dry and moist air at this point. It is important to note that the sensor positioned near the radial fan may indicate a relative humidity above 100\%, as its accuracy is only reliable up to 90\%\,\ac{rh}.

\subsection{Strain measurement}

The creep rack measures the sample strains over time with Digital Image Correlation \ac{dic}. For this purpose, the creep rack periodically records RGB images of the speckled specimen surfaces with a Grasshopper GS3-U3-51S5C camera equipped with a Kowa lens LM25JC5MC 2/3" 25mm with static focal length. During the recording, a motorized 2-axis positioning system moves the camera to predefined positions, capturing the samples' front and back sides at optimal working distance and sharpness. Recording both sides improves the measurement quality \citep{muller2015} and eliminates errors from out-of-plane motion. To enable two-sided measurement, the recorded surfaces are arranged perpendicular to the camera's viewing axis, and their speckling pattern is reflected into the camera by a mirror system 3D-printed with PETG. LEDs integrated into the mirrors provide a uniform, constant illumination of the sample surfaces. The light of these LEDs should not directly reflect into the mirrors to not decrease the image quality.

LabVIEW controls both the 2-axis positioning system and the image acquisition. The positioning system's control unit is a Phytron MCC-2 32-48 Mini Controller. It pilots two Sanyo Denki 103H7823-1740 stepping motors with a step angle of 1.8° and $1/256$ minimum step size. The movement axis controlling the camera's working distance has a thread pitch of \SI{3}{mm}. The other movement axis, driving between the samples, has a thread pitch of \SI{5}{mm}. Both stepping motors drive with $1/128$ steps, thus leading to a driving accuracy of 0.117 and 0.195\,\textmu m, respectively.

Despite the high accuracy, the experiments show that slight deviations in the steps and thermal expansion of the creep rack and the mirror supports can lead to minimal fluctuations in the camera's working distance. These fluctuations can lead to nonexistent volumetric strains measured by the \ac{dic}. Their strain magnitude of $\si{10^{-3}}$ to $\si{10^{-4}}$ is in a similar range to typical creep strain magnitudes. To correct for these, non-loaded speckled reference samples made from brass are connected to the sample holders. Their measured volumetric strains match these fictional strains and are used to correct the strain error of the wood samples. Early experiments of the creep rack did not yet employ such reference samples. In that case, the central horizontal strains of the shear samples with a shear plane normal in the longitudinal direction are used for the correction. They are considered suitable because that direction is very stiff, and the Arcan shape is designed to have nearly no physical uniaxial strains in the area between the V-notches. Later experiments showed proportionality between these strain components and the reference samples' strains, thus confirming their applicability for such strain correction. Another suggestion for alternative reference samples can be found in \citet{pan2013}.

\begin{figure}
    \centering
    \includegraphics[width=\textwidth]{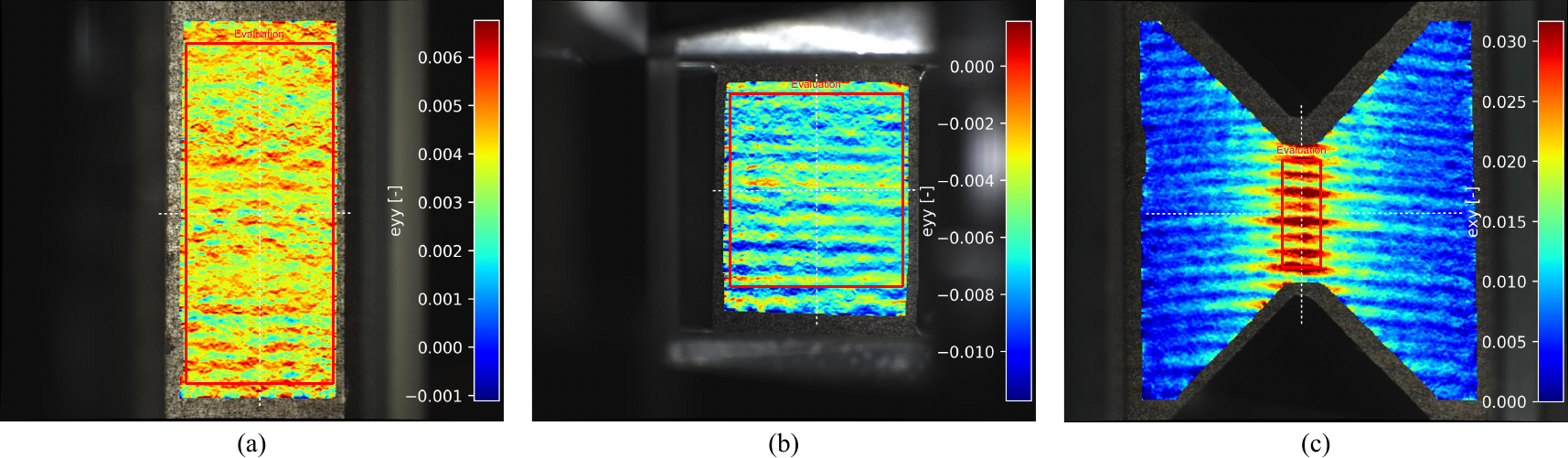}
    \caption{Region (red box) over which the strain is averaged for (a) tension samples, (b) compression samples, and (c) shear samples. The white dashed lines show the region's center axes. The exemplary fields are for (a) $\epsilon_{RR}^{t,R}$ for radial tension, (b) $\epsilon_{RR}^{c,R}$ for radial compression, and (c) $\epsilon_{TR}$ for TR shear plane. The region borders are based on the identified annual rings.}
    \label{fig_evaluation_roi}
\end{figure}

After every experiment, LabVIEW uploads the acquired photos and climate data to a research database that conforms to the FAIR data principles \citep{wilkinson2016}. The utilized research database, OpenBIS \citep{barillari2016}, stores the experimental data hierarchically, with links to calibration data, protocols, devices, and raw materials. 
An adapted version of the open-source 2D \acf{dic} Matlab program Ncorr \citep{blaber2015} is linked to that database and calculates the strains from the uploaded photos. Before correlating the images, Matlab's standard undistortion algorithm corrects the lens distortion from the image \citep{matlab2022}. The correction parameters result from a preceding camera calibration with a checkerboard pattern \citep{zhang2000, hanning2011}. The \ac{dic} subsequently determines the strain fields from the recorded speckle patterns over time. The strain fields are converted into scalar values by averaging them over a defined \ac{roi}. Fig.~\ref{fig_evaluation_roi} illustrates these regions for the compression, tension, and shear samples. When visible, the boundaries of the \ac{roi} are aligned with the annual rings of the specimen to ensure an equal ratio of earlywood and latewood in the averaged area. When no annual rings are observable, the \ac{roi} corresponds to the largest rectangular region that fits with a defined margin into the strain field.

\subsection{Data analysis}

The strain data obtained from the creep rack can close data gaps in the material parameters of existing rheological creep models of wood. Such hygromechanical creep models commonly utilize Maxwell or Kelvin-Voigt elements \citep{leicester1971, hanhijarvi1995, hanhijarvi2003, fortino2009, hassani2015, yu2022}. The equation to describe the time-dependent stiffness tensor $\boldsymbol{C}(t)$ of a generalized Maxwell-Wiechert model with $n$ elements, is \citep{gutierrez2014, hajikarimi2021}
\begin{align} \label{eq_maxwell}
    \boldsymbol{C}(t) = \boldsymbol{C}_\infty + \sum_{i=1}^n \left( \boldsymbol{C}_i \mathrm{e}^{  - t / \tau_i } \right) .
\end{align}
$\boldsymbol{C}_\infty$ is the stiffness tensor for time $t \rightarrow \infty$ and $\boldsymbol{C}_i$ and $\tau_i$ are the respective spring stiffness and characteristic time of the $i$-th element.
In contrast, the equation of the time-dependent elastic compliance tensor $\boldsymbol{C}^{-1}(t)$ of a generalized Kelvin-Voigt model with $n$ elements is \citep{gutierrez2014, hajikarimi2021}
\begin{align} \label{eq_kv}
    \boldsymbol{C}^{-1} (t) = \boldsymbol{C}_0^{-1} + \sum_{i=1}^n \boldsymbol{C}_i^{-1} \left( 1 - \mathrm{e}^{  - t / \tau_i } \right) .
\end{align}
$\boldsymbol{C}_0^{-1}$ is the initial elastic compliance tensor and $\boldsymbol{C}_i^{-1}$ and $\tau_i$ are the respective spring compliance and characteristic time of the $i$-th element.  

In the data analysis of the creep curve, we will utilize Kelvin-Voigt elements. Unlike Maxwell elements, they are serially superpositioned with the instantaneous elastic material response, allowing for easy extension with other rheological properties. Maxwell elements embed the immediate elasticity in the sum of the parallel strings. Therefore, the linear superposition of moisture-dependent rheological properties is not as straightforward. However, the strength of the Maxwell model lies in the description of relaxation and strain-recovery responses. In contrast, Kelvin-Voigt elements perform better for creep and stress-recovery responses \citep{gutierrez2014, hajikarimi2021}.

The parameters of the Kelvin-Voigt elements are commonly formulated relative to the initial elastic compliance. When applying ${C}_{i,jk}^{-1} = {C}_{0,jk}^{-1} \cdot \bar{g}^\mathrm{KV}_{i,jk}$ to Eq.\,\ref{eq_kv}, the compliance tensor component ${C}_{jk}^{-1}(t)$ takes the form \citep{gutierrez2014}
\begin{align} \label{eq_kv_g}
    {C}_{jk}^{-1}(t) = {C}_{0,jk}^{-1} \left( 1 + \sum_{i=1}^N \bar{g}^\mathrm{KV}_{i,jk} \left( 1 - \mathrm{e}^{ -t / \tau_i } \right) \right) .
\end{align}
The coefficients $\bar{g}^\mathrm{KV}_{i,jk}$ describe the characteristic compliances of the $i$-th Kelvin-Voigt element relative to the instantaneous elastic compliance for the tensor indices $jk$. The characteristic times of Eq.~\ref{eq_kv} have to cover the different representative periods of the experimental data. Additionally, the number of Kelvin-Voigt elements must neither produce under-fitting nor over-fitting \citep{hajikarimi2021}.
The experiment directly reveals the instantaneous elastic compliance. Therefore, only the relative characteristic compliances $\bar{g}^\mathrm{KV}_{i,jk}$ of the Kelvin-Voigt elements need to be determined. They get identified by a least-square fit over the experimental data. A Nelder-Mead algorithm from Python SciPy \citep{virtanen2020} solves the minimization problem.

\section{Results and discussion}

To show the creep rack performance, we conducted five viscoelastic creep experiments of Norway spruce at 65\% RH and $23 \pm 2$°C. Every experiment lasted \SI{30}{d} and comprised one compression test, one tension test, and one shear test per anatomical direction. Tab.\,\ref{tab_samples} documents the axial and lateral alignments of the measured sample surfaces and the applied loading degrees. The loading degrees vary between 19\% and 37\% to ensure linear creep \citep{schniewind1968, gressel1983}. We fitted four Kelvin-Voigt elements through all obtained creep curves to quantify the strain evolution over time. The Kelvin-Voigt elements' characteristic times are $\pmb{\tau} = \left[1, 10, 100, 1000\right]\si{h}$. These chosen times accomplish a good fitting quality and represent the various time scales during the experiment. $\pmb{\tau}$ is identical for all fitted creep curves, which is necessary for meaningfully relating the obtained creep parameters.

\begin{table}
\caption{Sample properties. The loading degree is relative to the ultimate axial strength, obtained from separate strength tests. t = tension, c = compression, s\,=\,shear loading type. R = radial, T = tangential, L = longitudinal direction.} \label{tab_samples}
\centering
\begin{tabular}{lcccccccccccc}
\textbf{Sample Type} & \textbf{tR} & \textbf{cR} & \textbf{tT} & \textbf{cT} & \textbf{tL} & \textbf{cL} & \textbf{sRT} & \textbf{sTR} & \textbf{sRL} & \textbf{sLR} & \textbf{sTL} & \textbf{sLT} \\
\midrule
\textbf{Loading type} & t & c & t & c & t & c & s & s & s & s & s & s \\
\textbf{Axial direction} & R & R & T & T & L & L & T & R & L & R & L & T \\
\textbf{Transverse direction} & L & L & R & R & T & T & R & T & R & L & T & L \\
\textbf{Axial Stress [$\boldsymbol{\mathrm{MPa}}$]} & $1.15$ & $-1.82$ & $0.86$ & $-1.79$ & $35.07$ & $-15.65$ & $0.51$ & $0.55$ & $1.89$ & $1.87$ & $1.55$ & $1.60$ \\
\textbf{Loading degree [$\boldsymbol{-}$]} & $0.19$ & $0.36$ & $0.19$ & $0.29$ & $0.25$ & $0.36$ & $0.26$ & $0.37$ & $0.35$ & $0.24$ & $0.22$ & $0.19$ \\
\end{tabular}
\end{table}

This section presents the obtained creep strain curves. We investigate the anisotropy between the anatomical directions, the loading direction asymmetries, and the time-dependent Poisson's ratios. The fitted Kelvin-Voigt elements quantify the creep curves. We determine each component of the creep compliance tensor by averaging the fitted creep curves of the corresponding strain direction, following Eqs. \ref{eq_creep_compliance_tensor_exp_axial}--\ref{eq_creep_compliance_tensor_exp_shear}. The asymmetric data from opposite loading directions over-determines the 9 independent components of the compliance tensor. Therefore, we symmetrize associated compression, tension, and shear data and identify the error of such symmetrization. Ultimately, we cluster the creep compliance tensor components into similar groups to reduce the number of independent variables. Through this, we recommend the minimum number of required experiments for sufficiently characterizing the creep orthotropy of Norway spruce.

\subsection{Kinematic observations}

\begin{figure}[!b]
    \centering
    \includegraphics[width=\textwidth]{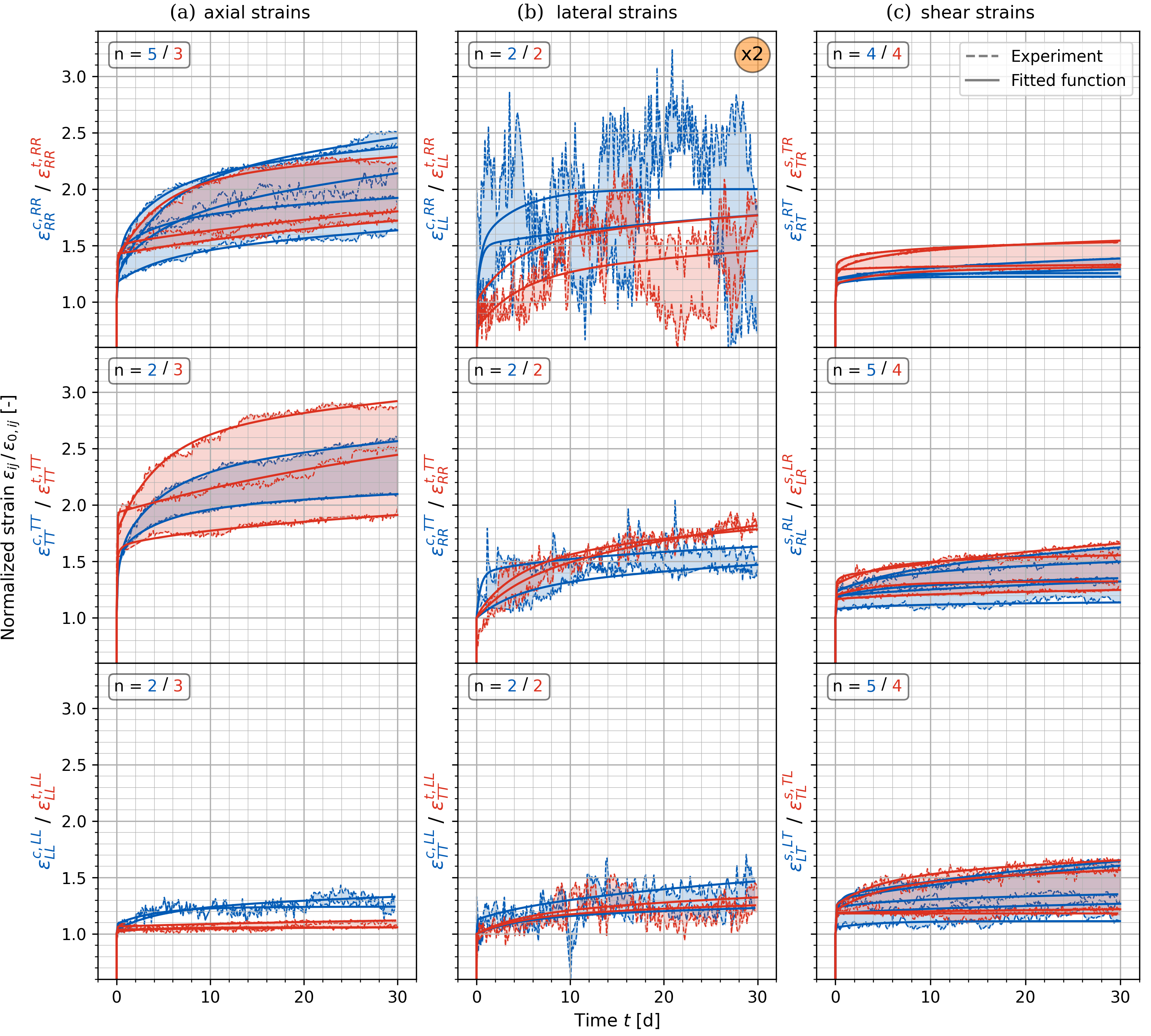}
    \caption{Raw normalized strain data of the creep experiment and their fitted Kelvin-Voigt series. $\epsilon_{jj}^{x,kk}$ corresponds to the strains of the uniaxial loading type $x$ (\textit{c} for compression, \textit{t} for tension), axial direction $kk$, and strain direction $jj$. $\epsilon_{jk}^{s,jk}$ represents the shear strains in the shear plane normal to $j$ induced by a force in direction $k$. (a) shows the axial strains for the uniaxially loaded samples, (b) the transverse strains of the uniaxially loaded samples, and (c) the shear strains of the shear samples. The strains $\epsilon_{LL}^{c,RR}$ and $\epsilon_{LL}^{t,RR}$ need to be multiplied by $2$.}
    \label{fig_results_strains}
\end{figure}

\begin{figure}[t]
    \centering
    \includegraphics[width=\textwidth]{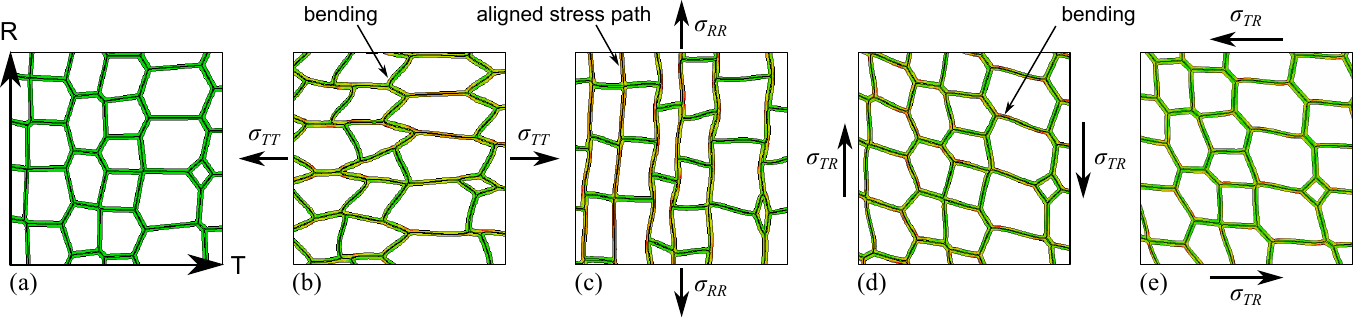}
    \caption{Deformations of the meso-scale structure (a) under the following load applications: (b) uniaxial stress $\sigma_{T}$, (c)\,uniaxial stress $\sigma_{R}$, (d) shear stress $\tau_{TR}$, and (e) shear stress $\tau_{RT}$. The colors represent qualitatively the von Mises stress.}
    \label{fig_structural_distortions}
\end{figure}

For comparability, the measured creep strains and the fitted Kelvin-Voigt models are normalized to the initial elastic strain as shown in Fig.\,\ref{fig_results_strains}. The data is grouped by anatomical direction and loading direction and comprises only those creep curves with sufficient data quality. $n$ indicates the sample count of each group. Fig.\,\ref{fig_results_strains}a presents the axial strains obtained from the uniaxial compression and tension tests. Here, the relative tangential creep is most pronounced, radial creep is slightly smaller, and longitudinal creep is smallest. These observations agree with \citet{cariou1987} and \cite{taniguchi2010}. We attribute those differences to the force transfer in the wood mesostructure. To rationalize such observations,  Fig.\,\ref{fig_structural_distortions}  shows the calculated deformations of an exemplary early-wood tissue, calculated with the Finite Element package ABAQUS. Fig.\,\ref{fig_structural_distortions}a illustrates the undeformed structure. Under tangential load (Fig.\,\ref{fig_structural_distortions}b), the forces activate the cell walls by bending close to the honeycomb's triple junctions. This bending leads to stress localizations and hinge formation with regionally higher loading degrees, potentially inducing local non-linear creep in hinges.
In contrast, the force paths under radial load in Fig.\,\ref{fig_structural_distortions}c are aligned with the tangential cell walls from the beginning, resulting in fewer stress localizations. Hence, radial creep is smaller than tangential creep. The serial combination of earlywood and latewood creep dominates the deformation. In comparison, longitudinal creep is considerably smaller because the longitudinal forces activate the entire cross-section of the cell walls. The uniform stress distribution circumvents stress localizations that would cause locally increased creep rates.
Whereas the loading direction seems to have little influence on tangential or radial creep, longitudinal compression is considerably larger than longitudinal tension, similar to \citet{gressel1983}. These findings do not entirely resemble \citet{ozyhar2013}, who also observed increased compression creep in radial and tangential directions. A potential reason for loading asymmetry in the longitudinal direction could be the wood fiber's tendency to buckle, reducing dimensional stability under compression.

Fig.\,\ref{fig_results_strains}b compares the lateral strains from uniaxial compression and tension tests. The lateral components $\epsilon_{RR}^{c, TT}$, $\epsilon_{RR}^{t, TT}$, $\epsilon_{TT}^{c, LL}$, and $\epsilon_{TT}^{t, LL}$ experience similar relative creep and no considerable loading direction asymmetry. Their magnitude is comparable to longitudinal axial creep. The lateral strains achieve a comparably low fitting quality with pronounced noise. This noise results from the small strain magnitude in lateral direction. This magnitude decreases with decreasing Poisson's ratio and thus the relative accuracy of \ac{dic}. For $\epsilon_{LL}^{c, RR}$ and $\epsilon_{LL}^{t, RR}$, where the Poisson's ratio is around $0.05$ \citep{niemz2023}, the lateral strains are so small that they are below the creep rack's measurement precision.

The shear creep strains, depicted in Fig.\,\ref{fig_results_strains}c, are all of similar magnitude. This similarity is counterintuitive because the shear stiffnesses and, thus, the deformation mechanisms vary considerably between the different shear directions. The shear creep magnitudes resemble $\epsilon_{RR}^{c, TT}$, $\epsilon_{RR}^{t, TT}$, $\epsilon_{TT}^{c, LL}$, and $\epsilon_{TT}^{t, LL}$. The strains $\epsilon_{TR}^{s, TR}$ and $\epsilon_{RT}^{s, RT}$ show a creep asymmetry, whereas the other shear planes behave rather symmetric. The asymmetry might derive from the mesostructure. $\epsilon_{TR}^{s, TR}$ in Fig.\,\ref{fig_structural_distortions}d is bending-dominated, leading to stress localizations with regionally higher creep. For $\epsilon_{RT}^{s, RT}$ in Fig.\,\ref{fig_structural_distortions}e, the bending of the cell walls is less pronounced due to the growth irregularity in the tangential direction. Hence, the stress localizations and maximum creep in the cell walls are smaller.

\begin{figure}[!t]
    \centering
    \includegraphics[width=0.8\textwidth]{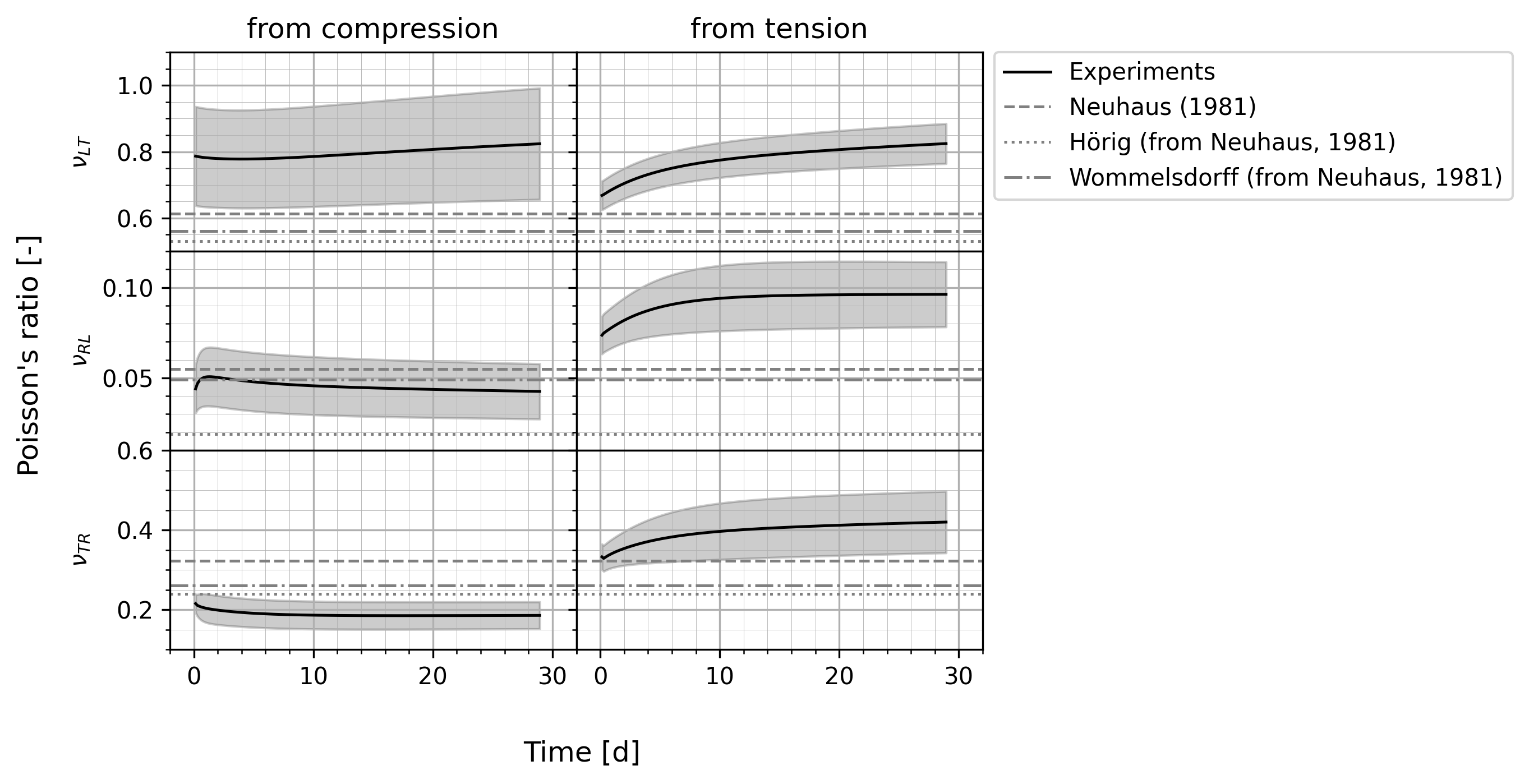}
    \caption{Development of the Poisson's ratio $\nu_{jk}$ over time for the axial direction $j$ and the lateral direction $k$. The solid line shows the averaged Poisson's ratio, whereas the gray area shows its standard deviation. The columns of the graphs indicate the loading type, and the rows the orientation of the Poisson's ratio. The Poisson's ratios for compression remain nearly constant over time, whereas for tension, they increase.}
    \label{fig_results_poisson}
\end{figure}

The elastic compliance's off-diagonal components are often expressed by their Poisson's ratio. Fig.\,\ref{fig_results_poisson} illustrates the Poisson ratios' development and standard deviation, calculated from the averaged axial and transverse strain fits. The ratios' order is similar to the elastic values from \citet{neuhaus1981}. Under compression, the Poisson's ratios remain nearly constant. Under tension, they increase. The observation for tension is consistent with \citet{ozyhar2013} but contradicts \citet{schniewind1972}, where the Poisson's ratios decrease over time. We observed that the determined Poisson's ratios are very sensitive to the fitting parameters of the axial and lateral strains. Even slight disproportions in these fitting parameters result in a time-dependent change of the Poisson's ratio. Early creep times are especially susceptible to this influence because the Kelvin-Voigt elements with low characteristic times $\tau_i$ are fitted over fewer data points. Due to this sensitivity on the fit quality and inconsistent trends for Poisson's ratio development in many publications \citep{hayashi1993, taniguchi2010, jiang2016}, we suspect Poisson's ratios are prone to errors. Thus, creep models should not use them as material parameters. \citet{hilton2001} supports this assumption and recommends directly quantifying the creep compliance's off-diagonal terms.

\subsection{Relative characteristic compliances}

\begin{figure}[t]
    \centering
    \includegraphics[width=\textwidth]{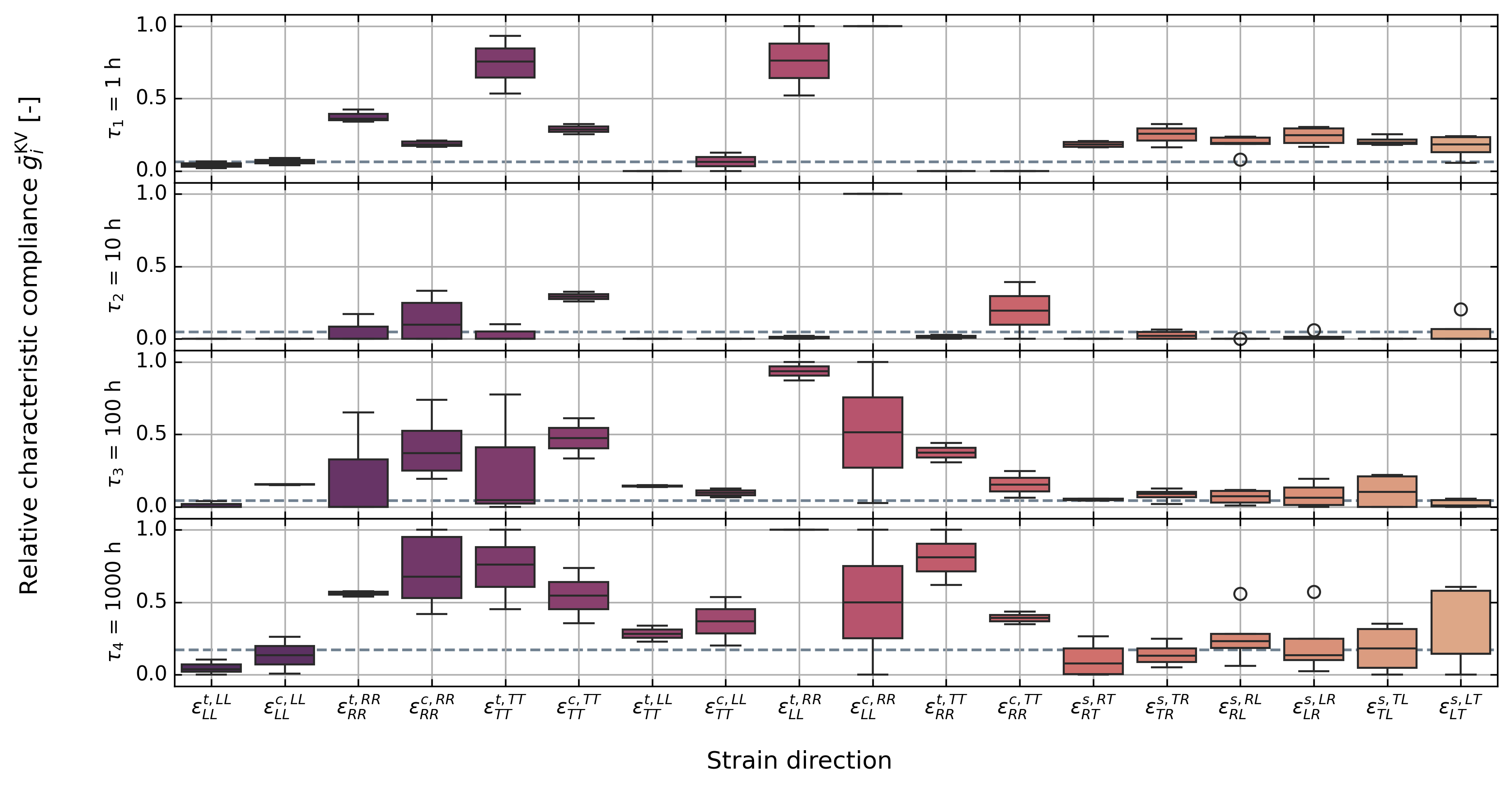}
    \caption{Comparison of the Kelvin-Voigt elements' relative characteristic compliance $g_{i}^\mathrm{KV}$ for each strain direction, grouped by characteristic time $\tau_i$. The strain direction $\epsilon_{jk}^{x,lm}$ corresponds to Fig.\,\ref{fig_results_strains}. The dashed line is the longitudinal creep data used in \citep{fortino2009}, fitted to $\boldsymbol{\tau} = [1, 10, 100, 1000]\,\mathrm{h}$.}
    \label{fig_results_comparison_coefficients}
\end{figure}

The Kelvin-Voigt elements' relative compliances enable the quantification of the creep strain relationships. Fig.\,\ref{fig_results_comparison_coefficients} shows the distribution of the relative characteristic compliances $\bar{g}^\mathrm{KV}_{i}$ for every strain component $\epsilon_{jk}^{x,lm}$ grouped by the corresponding characteristic times $\tau_i$. The dashed lines mark the characteristic compliances used by \citet{fortino2009}, adjusted to the figure's $\tau_i$, and are based on axial longitudinal creep at 12\% moisture content. These lines coincide well with the experimental relative characteristic compliances for axial longitudinal creep. It is interesting to observe that all shear creep components are similar in each group of $\tau_i$. The axial characteristic compliances in radial and tangential directions are higher than the ones for axial longitudinal and shear creep. The disparity between longitudinal compression and tension is slightly visible. The higher scatter for the lateral strains indicates the lower data quality. Furthermore, the magnitude of longitudinal creep and shear creep seems similar. All these insights resemble the direct observations from the strain data in Fig.\,\ref{fig_results_strains}.

Each strain direction's Kelvin-Voigt series describes one component of the time-dependent compliance tensor $\boldsymbol{C}^{-1} (t)$ from Eq.\,\ref{eq_creep_compliance_tensor}. Constructing $\boldsymbol{C}^{-1} (t)$ via Eq.\,\ref{eq_kv_g} requires an assembly of the relative characteristic compliances $g_i$ in a tensor:
\begin{equation} \label{eq_compliance_over_time}
\boldsymbol{C}^{-1} (t)
= \boldsymbol{C}_0^{-1} \left( 1 + \sum_{i=1}^N \bar{\boldsymbol{G}}_i^\mathrm{KV} \left( 1 - \mathrm{e}^{ -t / \tau_i } \right) \right) .
\end{equation}
$\boldsymbol{C}_0^{-1}$ is the initial compliance tensor, $\bar{\boldsymbol{G}}_i^\mathrm{KV}$ is a tensor with the direction-dependent relative characteristic compliances of the $i$-th Kelvin-Voigt elements, and $\tau_i$ are the corresponding characteristic times. $\bar{\boldsymbol{G}}_i^\mathrm{KV}$ follows the structure of Eqs.\,\ref{eq_creep_compliance_tensor_exp_axial}--\ref{eq_creep_compliance_tensor_exp_shear} and comprises the characteristic compliances as
\begin{align} \label{eq_characteristic_compliance_tensor}
\bar{\boldsymbol{G}}_i^\mathrm{KV} = 
\begin{bmatrix}
    \bar{g}^\mathrm{KV}_{i}( \epsilon_{RR}^{RR}) & \bar{g}^\mathrm{KV}_{i}( \epsilon_{RR}^{TT}) & \bar{g}^\mathrm{KV}_{i}( \epsilon_{LL}^{RR}) & 0 & 0 & 0 \\
     & \bar{g}^\mathrm{KV}_{i}( \epsilon_{TT}^{TT}) & \bar{g}^\mathrm{KV}_{i}( \epsilon_{TT}^{LL}) & 0 & 0 & 0 \\
     &  & \bar{g}^\mathrm{KV}_{i}( \epsilon_{LL}^{LL}) & 0 & 0 & 0 \\
     &  &  & \bar{g}^\mathrm{KV}_{i}( \epsilon_{RT}^{RT}) & 0 & 0 \\
     & \text{sym.}  &  &  & \bar{g}^\mathrm{KV}_{i}( \epsilon_{RL}^{RL}) & 0 \\
     &  &  &  &  & \bar{g}^\mathrm{KV}_{i}( \epsilon_{TL}^{TL}) \\
\end{bmatrix} ,
\end{align}
with
\begin{align}
\begin{aligned}
    &\min \left( \bar{g}^\mathrm{KV}_{i}(\epsilon_{jj}^{c,kk}), \bar{g}^\mathrm{KV}_{i}(\epsilon_{jj}^{t,kk})  \right) \leq \bar{g}^\mathrm{KV}_{i}(\epsilon_{jj}^{kk}) \leq \max \left( \bar{g}^\mathrm{KV}_{i}(\epsilon_{jj}^{c,kk}), \bar{g}^\mathrm{KV}_{i}(\epsilon_{jj}^{t,kk})  \right) &\text{ for all } i, j , \\
    &\min \left( \bar{g}^\mathrm{KV}_{i}(\epsilon_{jk}^{s,jk}), \bar{g}^\mathrm{KV}_{i}(\epsilon_{kj}^{s,kj})  \right) \leq \bar{g}^\mathrm{KV}_{i}(\epsilon_{jk}^{jk}) \leq \max \left( \bar{g}^\mathrm{KV}_{i}(\epsilon_{jk}^{s,jk}), \bar{g}^\mathrm{KV}_{i}(\epsilon_{jk}^{s,kj})  \right) &\text{ if } i \neq j .
\end{aligned}
\end{align}
$\bar{g}^\mathrm{KV}_{i}(\epsilon_{jk}^{x,lm})$ is the relative characteristic compliance of the $i$-th Kelvin-Voigt element, determined from the creep strain $\epsilon_{jk}^{lm}$ under loading type $x$ ($x$ equals $c$ for compression, $t$ for tension, $s$ for shear). The experimental data provides two values for each of the nine components $\bar{g}^\mathrm{KV}_{i}(\epsilon_{jk}^{lm})$ due to the recorded loading direction asymmetries.

\subsection{Scaling and data reduction}
Assembling the data in Eq.\,\ref{eq_characteristic_compliance_tensor} requires a data reduction, namely a symmetrization of compression, tension, and shear planes. One possible approach is to average associated directions, similar to \citet{bachtiar2017}:
\begin{align}
\begin{aligned}
    \bar{g}^\mathrm{KV}_{i}(\epsilon_{jj}^{kk}) &= \frac{ \bar{g}^\mathrm{KV}_{i}(\epsilon_{jj}^{c,kk}) + \bar{g}^\mathrm{KV}_{i}(\epsilon_{jj}^{t, kk}) }{2} &\text{ for all } i, j, \\
    \bar{g}^\mathrm{KV}_{i}(\epsilon_{jk}^{jk}) &= \frac{ \bar{g}^\mathrm{KV}_{i}(\epsilon_{jk}^{s,jk}) + \bar{g}^\mathrm{KV}_{i}(\epsilon_{kj}^{s, kj}) }{2} &\text{ if } i \neq j .
\end{aligned}
\end{align}
$\bar{g}^\mathrm{KV}_{i}(\epsilon_{jj}^{c,kk})$ and $\bar{g}^\mathrm{KV}_{i}(\epsilon_{jj}^{t, kk})$ match to the strain component $\epsilon_{jj}^{kk}$ from compression and tension, respectively, and $\bar{g}^\mathrm{KV}_{i}(\epsilon_{jk}^{s,jk})$ corresponds to the creep along shear plane $jk$ with $j, k \in \{ R, T, L \}$.
That leads to the symmetrized creep compliances in Eq.\,\ref{eq_appendix_prony_coefficients} in the Appendix.

Fig.\,\ref{fig_results_symmetrization}a depicts how the symmetrized creep compliance components, following Eq.\,\ref{eq_compliance_over_time}, develop over time. The half-transparent area marks how the creep compliance would appear if only data from tension or compression and one shear plane were used to construct the compliance. It is also interesting to compare the relative difference after \SI{30}{d} between the symmetrized creep compliance and the creep compliance that is purely assembled from tensile data and $\epsilon_{RT}^{s, RT}$, $\epsilon_{RL}^{s, RL}$, and $\epsilon_{TL}^{s, TL}$ data (see Fig.\,\ref{fig_results_symmetrization}b). The longitudinally loaded samples show again the most pronounced asymmetry, followed by the shear RT and TR planes and the radially loaded sample components.
The relative symmetrization error indicates the following:
\begin{enumerate}
    \item If a model incorporates a compliance matrix constructed from symmetrized values, the error is always below 8\%. That is comparatively small, considering the statistical variation of wood properties.
    \item If a model constructs the compliance tensor only from data in one loading direction, the error when simulating the opposite loading direction is up to 16\%. Thus, we advise constructing the creep compliances tensor from symmetrized values.
\end{enumerate}

\begin{figure}[thb]
    \centering
    \includegraphics[width=\textwidth]{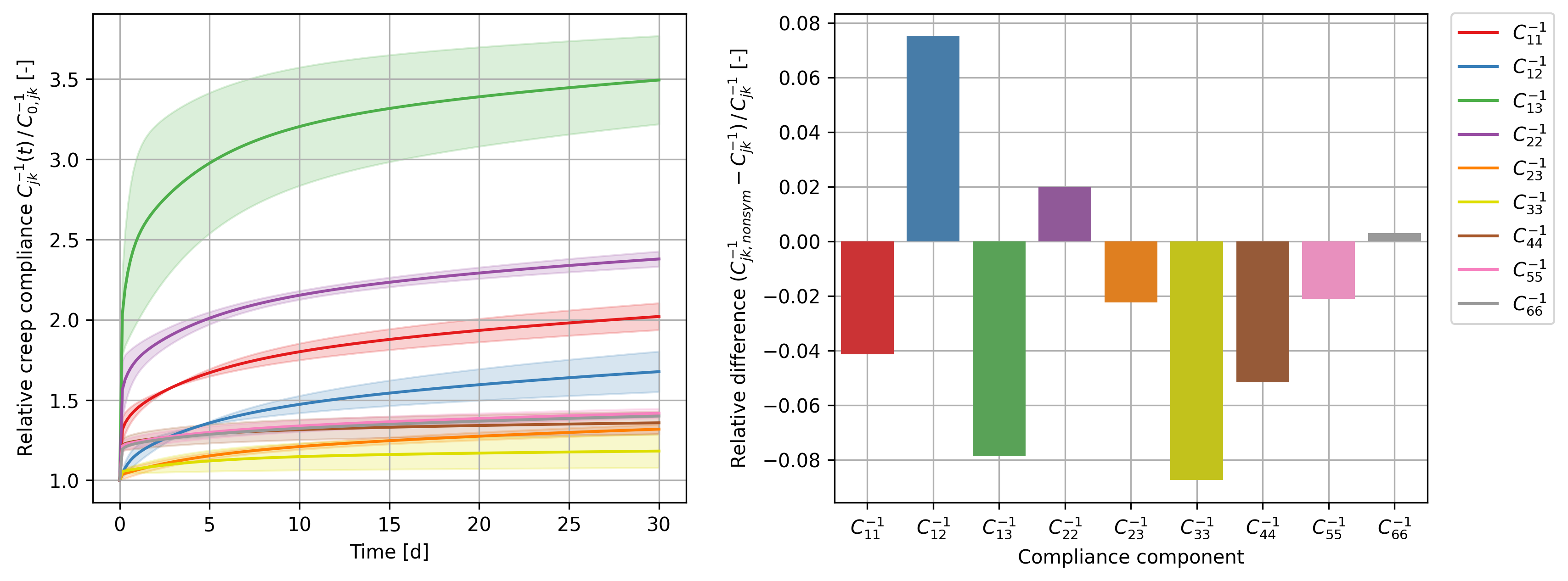}
    \caption{The solid lines in the left plot show the development of the symmetrized creep compliance components over time, normalized to the initial elastic compliance. The filled area marks the lower and upper bounds of the non-symmetrized compliances from only one loading type. The right plot shows the relative difference between the creep compliance $\boldsymbol{C}^{-1}_\mathrm{nonsym}(t)$ constructed from the tensile and the RT / RL / TL shear experiments at $t = \SI{30}{d}$ and the symmetrized creep compliance $\boldsymbol{C}^{-1}(t)$. A positive number indicates a higher compliance in tension than in compression.
}
    \label{fig_results_symmetrization}
\end{figure}

\begin{figure}[bt!]
    \centering
    \includegraphics[width=0.5\textwidth]{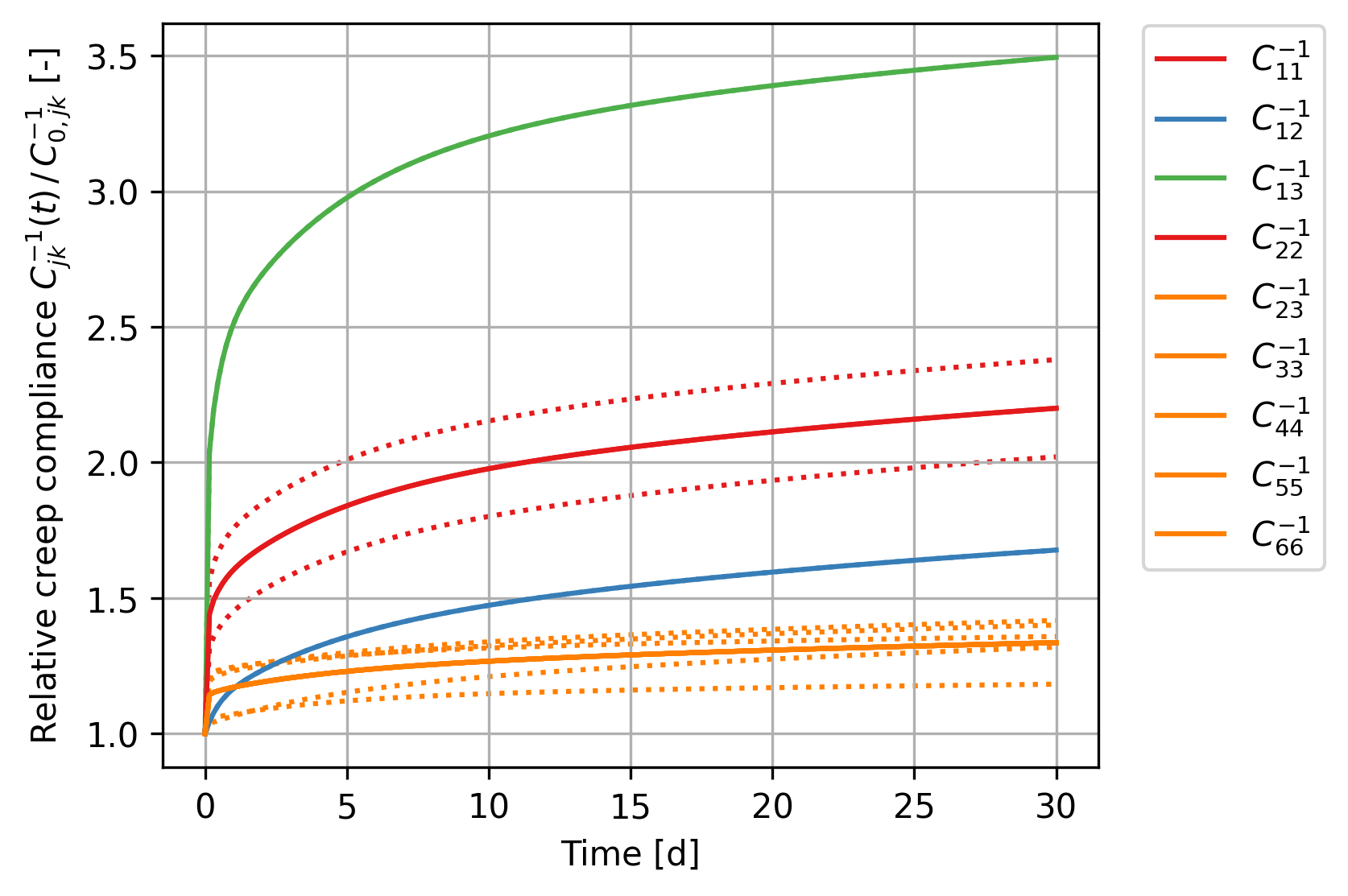}
    \caption{Grouping of the creep compliance components by k-means clustering. Each color represents one group. The dotted lines show the original curves before clustering.}
    \label{fig_results_clustering}
\end{figure}

Several creep models \citep{hanhijarvi2003, fortino2009, hassani2015, yu2022} use a proportionality assumption between the linear-elastic compliance tensor and the full creep compliance. Such scaling usually originates from data in one experimental direction, such as longitudinal creep from a 4-point bending test, what over-simplifies the anisotropy of wood creep \citep{frandsen2007}. Because we recorded creep in all anatomical directions, we don't need to perform such a simplification and can directly obtain every component of the time-dependent compliance matrix from the experiments. However, conducting such many experiments is connected with an extensive experimental effort. Therefore, it is rewarding to identify a set of proportionality factors that describe the relationship between elastic compliance and full creep compliance. Utilizing a k-means clustering algorithm from Python Scipy \citep{virtanen2020} one can bundle similar components of the creep compliance tensor into sets, leading to the grouping in Fig.\,\ref{fig_results_clustering}.
The Kelvin-Voigt element's relative characteristic compliances of these clusters are the following:
\begin{align}
\begin{aligned}
    k_1 &= \{ C^{-1}_{11}, C^{-1}_{22}\} &&:
    \quad \bar{g}^\mathrm{KV}_{1} = 0.398 \quad \bar{g}^\mathrm{KV}_{2} = 0.129 \quad \bar{g}^\mathrm{KV}_{3} = 0.345 \quad \bar{g}^\mathrm{KV}_{4} = 0.640 \ ,\\
    k_2 &= \{ C^{-1}_{12} \} &&:
    \quad \bar{g}^\mathrm{KV}_{1} = 0.000 \quad \bar{g}^\mathrm{KV}_{2} = 0.105 \quad \bar{g}^\mathrm{KV}_{3} = 0.264 \quad \bar{g}^\mathrm{KV}_{4} = 0.600 \ ,\\
    k_3 &= \{ C^{-1}_{13} \} &&:
    \quad \bar{g}^\mathrm{KV}_{1} = 0.880 \quad \bar{g}^\mathrm{KV}_{2} = 0.505 \quad \bar{g}^\mathrm{KV}_{3} = 0.725 \quad \bar{g}^\mathrm{KV}_{4} = 0.750 \ ,\\
    k_4 &= \{ C^{-1}_{23}, C^{-1}_{33}, C^{-1}_{44}, C^{-1}_{55}, C^{-1}_{66} \} &&:
    \quad \bar{g}^\mathrm{KV}_{1} = 0.140 \quad \bar{g}^\mathrm{KV}_{2} = 0.010 \quad \bar{g}^\mathrm{KV}_{3} = 0.081 \quad \bar{g}^\mathrm{KV}_{4} = 0.203 \ .
\end{aligned}
\end{align}

Obtaining the creep parameters for these four groups requires two uniaxial creep experiments that record axial and transverse strains: \textit{(i)} In the first test, the axial direction must be radial and the recorded lateral direction must be tangential. \textit{(ii)} In the second test, the axial direction must be longitudinal and the recorded lateral direction must be radial. Testing creep with these orientations provides the compliance components $C^{-1}_{11}$, $C^{-1}_{12}$, $C^{-1}_{33}$, and $C^{-1}_{13}$. Each component represents one of the clusters, enabling the full construction of the orthotropic compliance tensor. This approach can improve the quality of the creep compliance tensor tremendously compared to using only one proportionality factor. As a further simplification, the component $C^{-1}_{13}$ could be omitted additionally because the absolute elastic strain components in that direction are negligibly small, resulting in a marginal influence on a system's strain response.

\section{Conclusions}

We designed an automated climate-controlled creep rack that simultaneously measures wood creep in twelve samples. The samples comprise one dogbone tensile sample, one flat compression sample, and one Arcan shear sample for every anatomical direction and shear plane. Photo acquisition with Digital Image Correlation determines the sample's axial, lateral, and shear strains, parametrized by a fitted Kelvin-Voigt series. With this setup, each creep rack experiment provides experimental data for the complete viscoelastic or mechanosorptive creep compliance tensor while respecting compression-tension and shear asymmetries.

We utilized the creep rack to conduct viscoelastic experiments on Norway spruce at 65\%\,\ac{rh} and determined all nine independent components of the viscoelastic creep compliance tensor. The subsequent findings about these components' relationships provide the necessary knowledge for realistic creep model assumptions.
\begin{enumerate}
    \item Time-dependent Poisson's ratios are unsuitable as a material parameter for creep models due to a high susceptibility to measurement inaccuracies. The off-diagonal terms of the creep compliance matrix should be determined directly.
    \item The creep difference is up to 16\% between compression and tension and up to 10\% between complementary shear planes. The most pronounced loading direction asymmetry occurs for longitudinal and radial uniaxial loading and RT and TR shear loading.
    \item The creep compliance tensor is not fully proportional to the initial elastic compliance. The orthotropic tensor requires at least four independent parameters for its complete description. These parameters are obtainable from two creep experiments’ axial and lateral strain data. In the first test, the loading direction must be radial and the recorded lateral direction tangential. In the second test, the loading direction must be longitudinal and the recorded lateral direction radial.
\end{enumerate}

Scientists can improve their creep simulation models of wood based on two conclusions from this knowledge. Firstly, a model does not necessarily need to differentiate between opposite loading directions and complementary shear planes, which would imply considerable implementation effort. However, the creep compliance tensor should originate from data averaged from the asymmetric loading directions. Utilizing such a symmetrized compliance tensor induces a maximum error of moderate 8\% compared to an algorithm that differentiates asymmetric loading directions. Without symmetrization, that maximum error would be 16\%. Secondly, the state-of-the-art approach of uniformly scaling the viscoelastic compliance tensor to the initial elastic compliance tensor is invalid. Yet, a low number of creep experiments can adequately represent creep orthotropy. While the present publication provides Norway spruce's complete time-dependent compliance for one climate, the creep data for other species and humidities remains sparse. The assumption of obtaining the full creep compliance from two experiments can be transferred to other species to close data gaps with comparably low experimental effort.

Whereas 65\%\,\ac{rh} is a prevalent test climate, practical applications require creep data for changing moisture contents. Thus, future projects will utilize the creep rack to investigate Norway spruce's viscoelastic and mechanosorptive creep for multiple relative humidities. The data will provide insights into the relationships of various anatomical directions and climates. Ultimately, this will result in a complete moisture-dependent anisotropic rheological material database of Norway spruce. Such a reliable foundation will enable scientists to identify the essential interconnections for improving future hygromechanical models of wooden materials.

\section*{Acknowledgement}
The contribution to this work by Thomas Schnider for sample preparation and Alessia Ferrara for providing structural illustrations of wood's honeycomb structure, as well as the financial support from the Swiss National Science Foundation under SNF grant 200021\_192186 "Creep behavior of wood on multiple scales" is acknowledged. We thank Peter Niemz for fruitful discussions on the topic.







\section*{Appendix}

\setcounter{equation}{0}
\setcounter{figure}{0}
\setcounter{table}{0}

\renewcommand{\theequation}{A\arabic{equation}}
\renewcommand{\thefigure}{A\arabic{figure}}
\renewcommand{\thetable}{A\arabic{table}}

Eq.\,\ref{eq_appendix_prony_coefficients} shows the assembled relative characteristic compliances $\bar{\boldsymbol{G}}_{i}^\mathrm{KV}$ of the $i$-th Kelvin-Voigt elements that fulfill Eq.\,\ref{eq_compliance_over_time}. The data originates from the symmetrized loading directions. The corresponding characteristic times are $\tau_1 = \SI{1}{h}$, $\tau_2 = \SI{10}{h}$, $\tau_3 = \SI{100}{h}$, and $\tau_4 = \SI{1000}{h}$.

\begin{gather} \label{eq_appendix_prony_coefficients}
\begin{aligned}
\bar{\boldsymbol{G}}_{1}^\mathrm{KV} &=
\begin{bmatrix}
0.281 & 0.000 & 0.880 & 0.000 & 0.000 & 0.000 \\
      & 0.514 & 0.032 & 0.000 & 0.000 & 0.000 \\
      &       & 0.052 & 0.000 & 0.000 & 0.000 \\
      &       &       & 0.216 & 0.000 & 0.000 \\
      &\text{sym.}&   &       & 0.212 & 0.000 \\
      &       &       &       &       & 0.187 \\
\end{bmatrix}\\
\bar{\boldsymbol{G}}_{2}^\mathrm{KV} &=
\begin{bmatrix}
0.096 & 0.105 & 0.505 & 0.000 & 0.000 & 0.000 \\
      & 0.163 & 0.000 & 0.000 & 0.000 & 0.000 \\
      &       & 0.000 & 0.000 & 0.000 & 0.000 \\
      &       &       & 0.013 & 0.000 & 0.000 \\
      &\text{sym.}&   &       & 0.008 & 0.000 \\
      &       &       &       &       & 0.027 \\
\end{bmatrix}\\
\bar{\boldsymbol{G}}_{3}^\mathrm{KV} &=
\begin{bmatrix}
0.316 & 0.264 & 0.725 & 0.000 & 0.000 & 0.000 \\
      & 0.373 & 0.120 & 0.000 & 0.000 & 0.000 \\
      &       & 0.083 & 0.000 & 0.000 & 0.000 \\
      &       &       & 0.065 & 0.000 & 0.000 \\
      &\text{sym.}&   &       & 0.074 & 0.000 \\
      &       &       &       &       & 0.064 \\
\end{bmatrix}\\
\bar{\boldsymbol{G}}_{4}^\mathrm{KV} &=
\begin{bmatrix}
0.638 & 0.600 & 0.750 & 0.000 & 0.000 & 0.000 \\
      & 0.642 & 0.326 & 0.000 & 0.000 & 0.000 \\
      &       & 0.091 & 0.000 & 0.000 & 0.000 \\
      &       &       & 0.123 & 0.000 & 0.000 \\
      &\text{sym.}&   &       & 0.240 & 0.000 \\
      &       &       &       &       & 0.237 \\
\end{bmatrix}
\end{aligned}
\end{gather}

\bibliographystyle{humannat}
\bibliography{references}

\begin{thebibliography}{}

\bibitem[\protect\astroncite{Akter et~al.}{2023}]{akter2023}
Akter, S.~T., E.~Binder, and T.~K. Bader (2023).
\newblock Moisture and short-term time-dependent behavior of norway spruce
  clear wood under compression perpendicular to the grain and rolling shear.
\newblock {\em Wood Mater. Sci. Eng.}, 18(2): 580--593.

\bibitem[\protect\astroncite{Ando et~al.}{2023}]{ando2023}
Ando, K., R.~Nakamura, and T.~Kushino (2023).
\newblock Variation of shear creep properties of wood within a stem: Effects of
  macro- and microstructural variability.
\newblock {\em Wood Sci. Technol.}, 57(1): 93--110.

\bibitem[\protect\astroncite{Bachtiar et~al.}{2017}]{bachtiar2017}
Bachtiar, E.~V., S.~J. Sanabria, J.~P. Mittig, and P.~Niemz (2017).
\newblock Moisture-dependent elastic characteristics of walnut and cherry wood
  by means of mechanical and ultrasonic test incorporating three different
  ultrasound data evaluation techniques.
\newblock {\em Wood Sci. Technol.}, 51(1): 47--67.

\bibitem[\protect\astroncite{Barillari et~al.}{2016}]{barillari2016}
Barillari, C., D.~S.~M. Ottoz, J.~M. Fuentes-Serna, C.~Ramakrishnan, B.~Rinn,
  and F.~Rudolf (2016).
\newblock {{OpenBIS ELN-LIMS}}: {{An}} open-source database for academic
  laboratories.
\newblock {\em Bioinformatics}, 32(4): 638--640.

\bibitem[\protect\astroncite{Bengtsson et~al.}{2023}]{bengtsson2023}
Bengtsson, R., L.~Bergeron, R.~Afshar, M.~Mousavi, and E.~K. Gamstedt (2023).
\newblock Evaluating the viscoelastic shear properties of clear wood via
  off-axis compression testing and digital-image correlation.
\newblock {\em Mech. Time-Depend. Mater.}

\bibitem[\protect\astroncite{Blaber et~al.}{2015}]{blaber2015}
Blaber, J., B.~Adair, and A.~Antoniou (2015).
\newblock Ncorr: {{Open-Source 2D Digital Image Correlation Matlab Software}}.
\newblock {\em Exp. Mech.}, 55(6): 1105--1122.

\bibitem[\protect\astroncite{Cariou}{1987}]{cariou1987}
Cariou, J.-L. (1987).
\newblock {\em Caractérisation d'un matériau viscoélastique anisotrope : le
  bois}.
\newblock PhD thesis.
\newblock Talence, Université Bordeaux 1.

\bibitem[\protect\astroncite{Fortino et~al.}{2009}]{fortino2009}
Fortino, S., F.~Mirianon, and T.~Toratti (2009).
\newblock A {3D} moisture-stress {FEM} analysis for time dependent problems in
  timber structures.
\newblock {\em Mech. Time-Depend. Mater.}, 13(4): 333--356.

\bibitem[\protect\astroncite{Frandsen}{2007}]{frandsen2007}
Frandsen, H.~L. (2007).
\newblock {\em Selected Constitutive Models for Simulating the Hygromechanical
  Response of Wood}, number~10 in {{DCE Thesis}}.
\newblock Denmark: Department of Civil Engineering, Aalborg University.

\bibitem[\protect\astroncite{Gressel}{1983}]{gressel1983}
Gressel, P. (1983).
\newblock {Erfassung, systematische Auswertung und Erg{\"a}nzung bisheriger
  Untersuchungen {\"u}ber das rheologische Verhalten von Holz und
  Holzwerkstoffen - Ein Beitrag zur Verbesserung des
  Form{\"a}nderungsnachweises nach DIN 1052 "Holzbauwerke"}.
\newblock {Abschlussbericht} AIF 4289, AIF 5348, Versuchsanstalt f{\"u}r Stahl,
  Holz und Steine, Abt. Ingenieurholzbau, Universit{\"a}t Fridericiana
  Karlsruhe.

\bibitem[\protect\astroncite{{Gutierrez-Lemini}}{2014}]{gutierrez2014}
{Gutierrez-Lemini}, D. (2014).
\newblock {\em Engineering {{Viscoelasticity}}}.
\newblock {Boston, MA}: {Springer US}.

\bibitem[\protect\astroncite{Hajikarimi and
  Moghadas~Nejad}{2021}]{hajikarimi2021}
Hajikarimi, P. and F.~Moghadas~Nejad (2021).
\newblock {\em Applications of {{Viscoelasticity}}}.
\newblock {Elsevier}.

\bibitem[\protect\astroncite{Hanhijärvi}{1995}]{hanhijarvi1995}
Hanhijärvi, A. (1995).
\newblock Deformation kinetics based rheological model for the time-dependent
  and moisture induced deformation of wood.
\newblock {\em Wood Sci. Technol.}, 29(3): 191--199.

\bibitem[\protect\astroncite{Hanhijärvi and
  Mackenzie-Helnwein}{2003}]{hanhijarvi2003}
Hanhijärvi, A. and P.~Mackenzie-Helnwein (2003).
\newblock Computational {Analysis} of {Quality} {Reduction} during {Drying} of
  {Lumber} due to {Irrecoverable} {Deformation}. {I}: {Orthotropic}
  {Viscoelastic}-{Mechanosorptive}-{Plastic} {Material} {Model} for the
  {Transverse} {Plane} of {Wood}.
\newblock {\em J. Eng. Mech.}, 129(9): 996--1005.

\bibitem[\protect\astroncite{Hanning}{2011}]{hanning2011}
Hanning, T. (2011).
\newblock {\em {High Precision Camera Calibration}}.
\newblock {Wiesbaden}: {Vieweg+Teubner}.

\bibitem[\protect\astroncite{Hassani et~al.}{2015}]{hassani2015}
Hassani, M.~M., F.~K. Wittel, S.~Hering, and H.~J. Herrmann (2015).
\newblock Rheological model for wood.
\newblock {\em Comput. Methods Appl. Mech. Eng.}, 283(Supplement C):
  1032--1060.

\bibitem[\protect\astroncite{Hayashi et~al.}{1993}]{hayashi1993}
Hayashi, K., B.~Felix, and C.~Le~Govic (1993).
\newblock Wood viscoelastic compliance determination with special attention to
  measurement problems.
\newblock {\em Mater. Struct.}, 26(6): 370--376.

\bibitem[\protect\astroncite{Hering et~al.}{2012}]{hering2012_1}
Hering, S., S.~Saft, E.~Resch, P.~Niemz, and M.~Kaliske (2012).
\newblock Characterisation of moisture-dependent plasticity of beech wood and
  its application to a multi-surface plasticity model.
\newblock {\em Holzforschung}, 66(3).

\bibitem[\protect\astroncite{Hilton}{2001}]{hilton2001}
Hilton, H.~H. (2001).
\newblock Implications and {{Constraints}} of {{Time-Independent Poisson
  Ratios}} in {{Linear Isotropic}} and {{Anisotropic Viscoelasticity}}.
\newblock {\em J. Elast.}, 63(3): 221--251.

\bibitem[\protect\astroncite{Ho et~al.}{1993}]{ho1993}
Ho, H., M.~Y. Tsai, J.~Morton, and G.~L. Farley (1993).
\newblock A comparison of three popular test methods for determining the shear
  modulus of composite materials.
\newblock {\em Compos. Eng.}, 3(1): 69--81.

\bibitem[\protect\astroncite{Huč et~al.}{2018}]{huc2018}
Huč, S., S.~Svensson, and T.~Hozjan (2018).
\newblock Hygro-mechanical analysis of wood subjected to constant mechanical
  load and varying relative humidity.
\newblock {\em Holzforschung}, 72(10): 863--870.

\bibitem[\protect\astroncite{Jiang et~al.}{2016}]{jiang2016}
Jiang, J., B.~E. Valentine, J.~Lu, and P.~Niemz (2016).
\newblock Time dependence of the orthotropic compression {Young}’s moduli and
  {Poisson}’s ratios of {Chinese} fir wood.
\newblock {\em Holzforschung}, 70(11): 1093--1101.

\bibitem[\protect\astroncite{Kienzler and Schröder}{2019}]{kienzler2019}
Kienzler, R. and R.~Schröder (2019).
\newblock {\em Einführung in die {Höhere} {Festigkeitslehre}}.
\newblock Berlin, Heidelberg: Springer.

\bibitem[\protect\astroncite{Knigge and Schulz}{1966}]{knigge1966}
Knigge, W. and H.~Schulz (1966).
\newblock {\em {Grundriss der Forstbenutzung : Entstehung, Eigenschaften,
  Verwertung und Verwendung des Holzes und anderer Forstprodukte}}.
\newblock {Hamburg}: {Paul Parey}.

\bibitem[\protect\astroncite{Kollmann and Cote}{1968}]{kollmann1968}
Kollmann, F.~F. and W.~A. Cote (1968).
\newblock {\em Principles of {{Wood Science}} and {{Technology I Solid Wood}}},
  1 edition.
\newblock Springer-Verlag Berlin Heidelberg.

\bibitem[\protect\astroncite{Leicester}{1971}]{leicester1971}
Leicester, R.~H. (1971).
\newblock A rheological model for mechano-sorptive deflections of beams.
\newblock {\em Wood Sci. Technol.}, 5(3): 211--220.

\bibitem[\protect\astroncite{Liu}{1993}]{liu1993}
Liu, T. (1993).
\newblock Creep of wood under a large span of loads in constant and varying
  environments.
\newblock {\em Holz Roh- Werkst.}, 51(6): 400--405.

\bibitem[\protect\astroncite{Morlier}{1994}]{morlier1994}
Morlier, P., ed. (1994).
\newblock {\em Creep in {{Timber Structures}}}.
\newblock London: CRC Press.

\bibitem[\protect\astroncite{Müller et~al.}{2015}]{muller2015}
Müller, U., A.~Ringhofer, R.~Brandner, and G.~Schickhofer (2015).
\newblock Homogeneous shear stress field of wood in an {{Arcan}} shear test
  configuration measured by means of electronic speckle pattern interferometry:
  Description of the test setup.
\newblock {\em Wood Sci. Technol.}, 49(6): 1123--1136.

\bibitem[\protect\astroncite{Neuhaus}{1981}]{neuhaus1981}
Neuhaus, F.-H. (1981).
\newblock {\em {Elastizit{\"a}tszahlen von Fichtenholz in Abh{\"a}ngigkeit von
  der Holzfeuchtigkeit}}.
\newblock PhD thesis.
\newblock Inst. f{\"u}r Konstruktiven Ingenieurbau, Ruhr-Universit{\"a}t
  Bochum.

\bibitem[\protect\astroncite{Niemz et~al.}{2023}]{niemz2023}
Niemz, P., A.~Teischinger, and D.~Sandberg, eds. (2023).
\newblock {\em Springer {{Handbook}} of {{Wood Science}} and {{Technology}}},
  Springer {{Handbooks}}.
\newblock Cham: Springer International Publishing.

\bibitem[\protect\astroncite{Ozyhar et~al.}{2013}]{ozyhar2013}
Ozyhar, T., S.~Hering, and P.~Niemz (2013).
\newblock Viscoelastic characterization of wood: {{Time}} dependence of the
  orthotropic compliance in tension and compression.
\newblock {\em J. Rheol.}, 57(2): 699--717.

\bibitem[\protect\astroncite{Pan et~al.}{2013}]{pan2013}
Pan, B., L.~Yu, D.~Wu, and L.~Tang (2013).
\newblock Systematic errors in two-dimensional digital image correlation due to
  lens distortion.
\newblock {\em Opt. Lasers Eng.}, 51(2): 140--147.

\bibitem[\protect\astroncite{Pierron and Vautrin}{1996}]{pierron1996}
Pierron, F. and A.~Vautrin (1996).
\newblock The 10 {$^\circ$} off-axis tensile test: {{A}} critical approach.
\newblock {\em Compos. Sci. Technol.}, 56(4): 483--488.

\bibitem[\protect\astroncite{Ranta-Maunus}{1975}]{rantamaunus1975}
Ranta-Maunus, A. (1975).
\newblock The viscoelasticity of wood at varying moisture content.
\newblock {\em Wood Sci. Technol.}, 9(3): 189--205.

\bibitem[\protect\astroncite{Schmidt and Kaliske}{2006}]{schmidt2006}
Schmidt, J. and M.~Kaliske (2006).
\newblock {Zur dreidimensionalen Materialmodellierung von Fichtenholz mittels
  eines Mehrfl{\"a}chen-Plastizit{\"a}tsmodells}.
\newblock {\em Holz Roh- Werkst.}, 64(5): 393--402.

\bibitem[\protect\astroncite{Schniewind}{1968}]{schniewind1968}
Schniewind, A.~P. (1968).
\newblock Recent progress in the study of the rheology of wood.
\newblock {\em Wood Sci. Technol.}, 2(3): 188--206.

\bibitem[\protect\astroncite{Schniewind and Barrett}{1972}]{schniewind1972}
Schniewind, A.~P. and J.~D. Barrett (1972).
\newblock Wood as a linear orthotropic viscoelastic material.
\newblock {\em Wood Sci. Technol.}, 6(1): 43--57.

\bibitem[\protect\astroncite{Shimazaki and Ando}{2024}]{shimazaki2024}
Shimazaki, K. and K.~Ando (2024).
\newblock Analysis of shear creep properties of wood via modified {{Burger}}
  models and off-axis compression test method.
\newblock {\em Wood Sci. Technol.}, 58(4): 1473--1490.

\bibitem[\protect\astroncite{Taniguchi et~al.}{2010}]{taniguchi2010}
Taniguchi, Y., K.~Ando, and H.~Yamamoto (2010).
\newblock Determination of three-dimensional viscoelastic compliance in wood by
  tensile creep test.
\newblock {\em J. Wood Sci.}, 56(1): 82--84.

\bibitem[\protect\astroncite{{The MathWorks Inc.}}{2022}]{matlab2022}
{The MathWorks Inc.} (2022).
\newblock {\em MATLAB version: 9.13.0 (R2022b)}.
\newblock Natick, Massachusetts, United States: The MathWorks Inc.

\bibitem[\protect\astroncite{Tong et~al.}{2020}]{tong2020}
Tong, D., S.~A. Brown, D.~Corr, and G.~Cusatis (2020).
\newblock Wood creep data collection and unbiased parameter identification of
  compliance functions.
\newblock {\em Holzforschung}, 74(11): 1011--1020.

\bibitem[\protect\astroncite{Toratti}{1992}]{toratti1992}
Toratti, T. (1992).
\newblock Modelling the creep of timber beams.
\newblock {\em J. Struct. Mech.}, 25(1): 12--35.

\bibitem[\protect\astroncite{Toratti and Svensson}{2000}]{toratti2000}
Toratti, T. and S.~Svensson (2000).
\newblock Mechano-sorptive experiments perpendicular to grain under tensile and
  compressive loads.
\newblock {\em Wood Sci. Technol.}, 34(4): 317--326.

\bibitem[\protect\astroncite{Trendelenburg}{1955}]{trendelenburg1955}
Trendelenburg, R. (1955).
\newblock {\em {Das Holz als Rohstoff}}, 2 edition.
\newblock {Munich, Germany}: {Carl Hanser Verlag GmbH \& Co. KG}.

\bibitem[\protect\astroncite{Virtanen et~al.}{2020}]{virtanen2020}
Virtanen, P., R.~Gommers, T.~E. Oliphant, M.~Haberland, T.~Reddy,
  D.~Cournapeau, E.~Burovski, P.~Peterson, W.~Weckesser, J.~Bright, et~al.
  (2020).
\newblock {{SciPy} 1.0: Fundamental Algorithms for Scientific Computing in
  Python}.
\newblock {\em Nat. Methods}, 17: 261--272.

\bibitem[\protect\astroncite{Wilkinson et~al.}{2016}]{wilkinson2016}
Wilkinson, M.~D., M.~Dumontier, I.~J. Aalbersberg, G.~Appleton, M.~Axton,
  A.~Baak, N.~Blomberg, J.-W. Boiten, L.~B. da~Silva~Santos, P.~E. Bourne,
  et~al. (2016).
\newblock The {FAIR} {Guiding} {Principles} for scientific data management and
  stewardship.
\newblock {\em Sci Data}, 3(1): 160018.

\bibitem[\protect\astroncite{Xavier et~al.}{2004}]{xavier2004}
Xavier, J., N.~Garrido, M.~Oliveira, J.~Morais, P.~Camanho, and F.~Pierron
  (2004).
\newblock A comparison between the {{Iosipescu}} and off-axis shear test
  methods for the characterization of {{Pinus Pinaster Ait}}.
\newblock {\em Compos. Pt. A-Appl. Sci. Manuf.}, 35(7-8): 827--840.

\bibitem[\protect\astroncite{Yu et~al.}{2022}]{yu2022}
Yu, T., A.~Khaloian, and J.-W. Van De~Kuilen (2022).
\newblock An improved model for the time-dependent material response of wood
  under mechanical loading and varying humidity conditions.
\newblock {\em Eng. Struct.}, 259: 114116.

\bibitem[\protect\astroncite{Zhang}{2000}]{zhang2000}
Zhang, Z. (2000).
\newblock A flexible new technique for camera calibration.
\newblock {\em IEEE Trans. Pattern Anal. Mach. Intell.}, 22(11): 1330--1334.

\bibitem[\protect\astroncite{ZwickRoell}{2023}]{zwickroell2023}
ZwickRoell (2023).
\newblock Kappa {Multistation} creep testing machine for plastics testing.
\newblock Product {Information} 88\_959\_ZRF\_06.2023, ZwickRoell GmbH \& Co.
  KG, Ulm, Baden-Würtemberg, Germany.

\end{thebibliography}
\end{document}